\documentclass[aps,prb,twocolumn,showpacs,floatfix]{revtex4}
\usepackage{amssymb}
\usepackage{amsmath}
\usepackage{dsfont}
\usepackage{graphicx}
\usepackage{dcolumn}
\usepackage{bm}
\usepackage{color}
\usepackage[colorlinks=true,linkcolor=blue,citecolor=blue,filecolor=blue,pagecolor=blue,urlcolor=blue,bookmarks=true,bookmarksopen=true,bookmarksopenlevel=3,plainpages=false,pdfpagelabels=true]{hyperref}
\newcommand{\PRL}{{ Phys. Rev. Lett. }}
\newcommand{\PRB}{{ Phys. Rev. B }}

\newcommand{\PR}{{ Phys. Rev. }}
\newcommand{\ZP}{{ Z.Phys. }}
\newcommand{\RMP}{{ Rev. Mod. Phys. }}

\newcommand{\APL}{{ Appl. Phys. Lett. }}
\newcommand{\EPL}{{ Eur. Phys. Lett. }}
\begin{document}
\title{Scattering approach to frequency-dependent current noise in Fabry-P\'erot graphene devices} 
 \author{Jan Hammer}
 \author{Wolfgang Belzig}
 \affiliation{Fachbereich Physik, Universit\"at Konstanz, 78457 Konstanz, Germany}
 \date{\today}
\begin{abstract}
We study finite-frequency quantum noise and photon-assisted electron transport 
through a wide and ballistic graphene sheet sandwiched between two metallic leads. 
The elementary excitations allow as to examine the differences between effects related 
to Fabry-P\'erot like interferences and signatures caused by correlations of coherently 
scattered particles in electron- and hole-like parts of the Dirac spectrum. 
We identify different features in the current-current 
auto- and cross-correlation spectra and trace them back to the interference patterns 
of the product of transmission- and reflection amplitudes which define the integrands 
of the involved correlators. At positive frequencies the correlator of the 
auto-terminal noise spectrum with final- and initial state associated to the measurement 
terminal is dominant. Phase jumps occur within the interference patterns of corresponding integrands,
which also reveal the intrinsic energy scale of the two-terminal graphene setup. 
The excess noise spectra, as well as the cross-correlation ones, 
show large fluctuations between positive and negative values. 
Oscillatory signatures of the cross-correlation noise are due to an alternating behavior of the integrands.
\end{abstract}
\maketitle
\section{Introduction} 
Ballistic electron transport~\cite{Wharam88,vanWees88} in two-terminal graphene systems is in the focus
of intensive studies ever since the pioneering experiments on
single-layer carbon.~\cite{CastroNeto06,Geim07} 
The Dirac Hamiltonian~\cite{CastroNeto06,Dirac28}
describes charge transport close to the charge-neutrality point and
leads to a linear dispersion relation $\epsilon=\hbar v_F k$. This
allows to observe several relativistic phenomena in solid-state
system, such as Klein tunneling~\cite{Klein28,Dombey99,Katsnelson07,Shytov08,Young09,Sonin09}
or the Zitterbewegung.~\cite{Katsnelson06,Schliemann08b,Rusin07b}  
In the very early works on graphene the minimal
conductivity~\cite{Novoselov05,Zhang05,Katsnelson06,Katsnelson06a,Ziegler07}
$G\approx e^2/h$ per valley and pseudo-spin at the charge-neutrality
point has been found and stimulated the research on current and noise
properties. The current-current correlations around the minimal
conductivity lead to a zero frequency sub-Poissonion Fano factor with
a maximal value of $F=1/3$,~\cite{Tworzydlo06,ZhuGuo07,Groth08,Danneau08,DiCarlo08} 
remarkably similar to diffusive systems as disordered metals.~\cite{Beenakker92,Nagaev92,Lewenkopf08}  
The suppression of the Fano factor below the Poissonian value originates from noiseless,
open quantum channels that are found at all conductance minima in
graphene-based two-terminal structures~\cite{ZhuGuo07}, and can be
explained as an interplay between Klein tunneling, resonant tunneling
and pseudo-spin matching. This pseudo-diffusive behavior~\cite{Rycerz09} is due to
the special band-structure of graphene.  Without impurity scattering,
coherent transport through such a graphene sheet~\cite{Sonin08} gives
rise to the same shot noise as in classical diffusive systems. 
The opening of a gap~\cite{LopezRodriguez08} in the quasiparticle spectrum
leads to an enhanced Fano factor.~\cite{ZhuGuo07}
Such a gap can be opened for example by photon-assisted tunneling, as
shown recently for the case of a graphene p-n
junction~\cite{Williams07} with a linear potential drop across the
interface.~\cite{Fistul07,Syzranov08} 
There, Landau-Zener like transitions stimulated by photon emission or absorption via resonant
interaction of propagating quasiparticles in graphene with an
irradiating electric field lead to hopping between different trajectories.\\
\begin{figure}[tb]
	\centering
	\includegraphics[width=0.8\columnwidth]{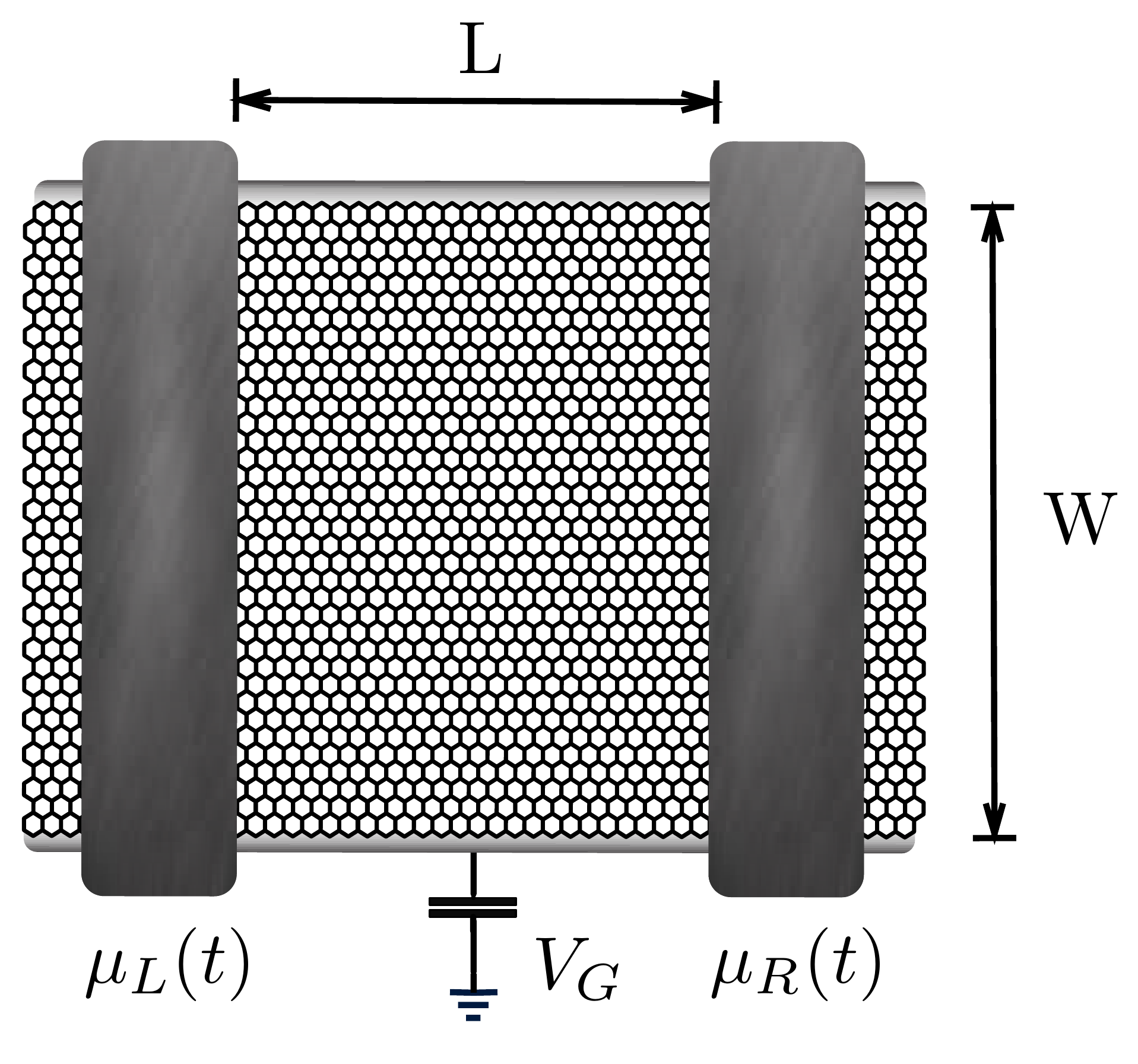}\\
	\caption{Wide graphene strip ($W\gg L$) sandwiched between two heavily-doped, 
			metallic graphene leads. The Fermi-level of the sheet 
			can be tuned by a center gate voltage $V_g$. 
			Electron- and hole states are injected via the time-dependent 
			bias-voltages in left- and right leads $\mu_{L/R}(t)$.}
	\label{fig1}
\end{figure}
The scattering approach as put forward by Landauer and
B\"uttiker~\cite{BlanterButtiker00} has been applied to ac-driven
charge transport~\cite{TienGordon62,Grifoni98,Platero04,Kohler05} through a metal-graphene interface with an abrupt
potential change.~\cite{Trauzettel07} The metal can be formed by a
graphene lead strongly electrostatically doped by a gate potential, thus
shifting the Dirac point far away from the Fermi energy.  In this
work we adopt the formalism and parameterization
introduced in Refs.~\cite{Beenakker06,Trauzettel07} and calculate the
finite-frequency current-current correlations at finite dc- and
ac-bias voltages in the system depicted in Fig.~\ref{fig1}. 
We complement recent results on ac-transport in
Fabry-P\'erot graphene devices of Ref.~\cite{Rocha09}, in which the
influence of different boundary conditions, i.e. zigzag or armchair
configurations, on the Fabry-P\'erot patterns in a combined
Tien-Gordon/tight-binding approach has been investigated. 
The influence on transmission properties of a time-dependent potential barrier in 
a graphene monolayer has been investigated.~\cite{Zeb08} 
In our work the transverse boundary effects
are central and we assume so-called infinite mass
boundaries~\cite{Tworzydlo06,Berry87} describing a
short but wide ($L\ll W$) graphene strip. 
We focus on the interplay between the Dirac-spectrum with the Fabry-Perot interferences.

Interestingly, the well-known oscillations as function of gate voltage on a
scale of the return frequency $\hbar v_F/L$, related to the length $L$ of the
graphene sheet, can be seen as a reminiscence of Zitterbewegung.~\cite{Katsnelson06} 
The role of the complex reflection amplitude and the onset of contributions of
scattering states coming from terminal $\alpha$ and being scattered
into terminal $\beta$ will be the key characteristics in our
discussion of the results for the noise as function of bias
voltage and frequency. As a consequence of these onsets the
oscillations add up de- or constructively depending on
the precise values of voltage and frequency. In our setup, 
the separation of oscillations caused by the
Fabry-P\'erot reflections and effects caused by the band-structure of
the Dirac Hamiltonian is a priori not obvious. In both cases
phase-coherent transport is essential. However, for charge injection
either into the conduction or the valence band only, effects like
Zitterbewegung should not be present and all oscillating features of
the noise spectra have to be of Fabry-P\'erot nature.
\section{Dirac equation and scattering formalism} 
The ballistic graphene~\cite{Miao07,Laakso08,Schuessler09} sheet considered in the following can be
described by the two-dimensional Dirac equation for the two-component
spinor $\hat{\Psi}=(\hat{\Psi}_1, \hat{\Psi}_2)^T$ with indices
referring to the two pseudo-spins of the carbon sub-lattices. 
Throughout this work we will neglect inter-valley scattering and Coulomb interactions. 
We only consider the interaction of the electrons with the radiation field in the form of
photon-assisted transitions. With Fermi velocity $v_F$ the Dirac
equation can be cast into the form
\begin{align}
	&	\left[-i v_F \hbar 	
			\left(\begin{array}{cc}
				0 & \partial_x - i \partial_y\\
				\partial_x + i \partial_y & 0
			\end{array}\right)
		-\mu({\mathbf x},t)\right]{\hat{\Psi}}({\mathbf{x}},t) \nonumber \\
	&	= i \hbar \partial_t {\hat{\Psi}}({\mathbf{x}},t) \,.
\label{diraceq}
\end{align}
The electrochemical potential $\mu({\mathbf{x}},t)$
includes static and harmonically driven potentials in the leads plus a
static gate voltage in the graphene sheet.
\begin{align}
	\mu({\mathbf x},t)=\left\{
	\begin{array}{ccc}
		\mu_L + eV_{\mathrm{ac,L}} \cos(\omega t) & \mathrm{ if } & x<0\\
		eV_g & \mathrm{ if } & 0<x<L\\
		\mu_R + eV_{\mathrm{ac,R}}\cos(\omega t) & \mathrm{ if } & x>L
	\end{array}\right..
	\label{potentials}
\end{align}
Making use of the Tien-Gordon ansatz, we write the solution to the
time-dependent Dirac equation as a sum over
PAT modes:
\begin{align}
	{\hat{\Psi}}({\mathbf{x}},t)   =  &  {\hat{\Psi}}_{\mathrm{0}}({\mathbf{x}},t) e^{-i(eV_{ac}/\hbar \omega)\sin(\omega t)} \\
	  = & \sum\limits_{m=-\infty}^{\infty} J_m \left(\frac{eV_{ac}}{\hbar \omega } \right) {\hat{\Psi}}_{\mathrm{0}}({\mathbf{x}},t) e^{-i m\omega t}\\
	\mathrm{ where }  \qquad &  {\hat{\Psi}}_{\mathrm{0}}({\mathbf{x}},t)={\hat{\Psi}}_{\mathrm{0}}({\mathbf{x}}) e^{- i \epsilon t}
	\label{states}
\end{align}
The advantage of this ansatz is that the scattering problem has to be
solved for the time-independent case only. Therefor, in 
terminals $\gamma=L,R$ we define stationary solutions 
${\hat{\Psi}}_{\mathrm{0}}({\bf{x}},t)={\hat{\Psi}}_{\epsilon}({\bf{x}})
e^{- i \epsilon t}$ by the equation
\begin{align}&
\left[-i v_F \hbar 
\left(\begin{array}{cc}
0 & \partial_x - i \partial_y\\
\partial_x + i \partial_y & 0
\end{array}\right)
-\mu_{\gamma}\right]{\hat{\Psi}}_{\mathrm{0}}({\mathbf{x}})  \\
&= \epsilon {\hat{\Psi}}_{\mathrm{0}}({\mathbf{x}}) \,.
\label{diraceq2}
\end{align}
\begin{figure}[tb]
	\centering
		\includegraphics[width=0.6\columnwidth]{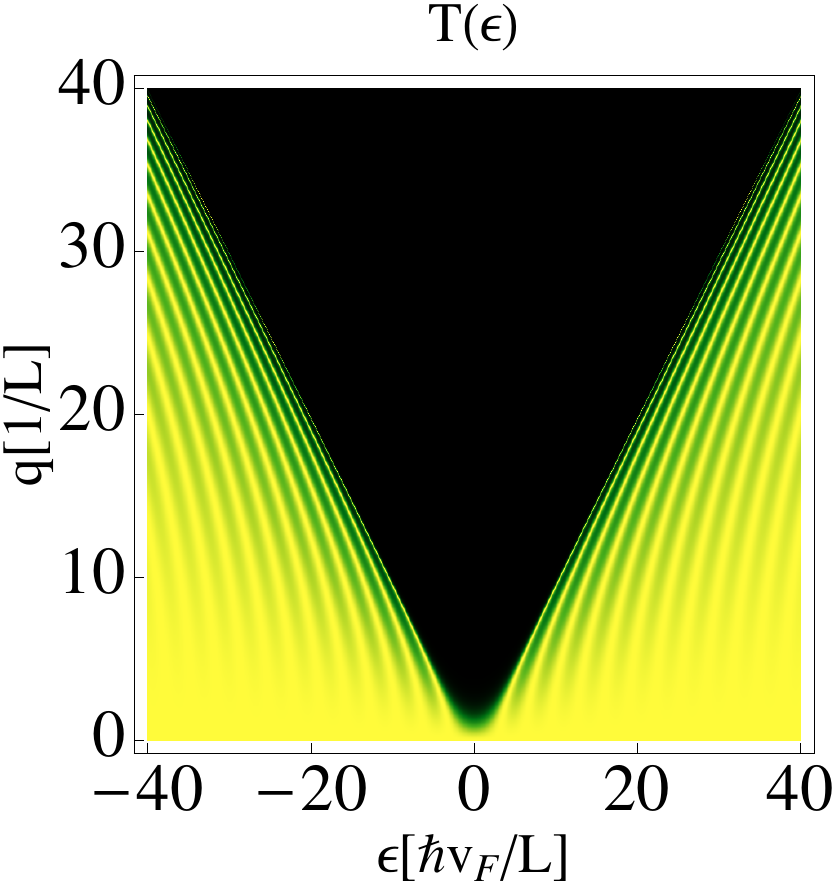}
		\caption{(color online) Transmission probability $T(\epsilon)=\left|t(\epsilon) \right|^2$ 
		as a function of energy and transverse momentum q. 
		}
	\label{fig2}
\end{figure}
The basis states in graphene
can be constructed as a superposition of left- and right movers,
\begin{equation}
\hat{\Psi}_0({\mathbf x})= \sum\limits_{k,q}^{} \left[{{\Psi}}_{\mathrm{0,+}}^{k,q} \hat{a}_{k,q} + {{\Psi}}_{\mathrm{0,-}}^{k,q} \hat{a}_{-k,q} \right] \, .
\end{equation}
$\alpha(\epsilon)$ describes the angle 
between the momentum of a quasiparticle
and it's y-component $q$ in region $x=0 \ldots L$ of the graphene sheet. 
Then the pseudo-spinors can be parametrized as
\begin{align}
{{\Psi}}_{\mathrm{0,+}}^{k,q} =&\frac{e^{i q y +i k(\epsilon)x}}{\sqrt{\cos\alpha(\epsilon)}}
	\left(\begin{array}{c}
		e^{-i \alpha(\epsilon)/2} \\
		e^{i \alpha(\epsilon)/2}
		\end{array}\right)
\\
{{\Psi}}_{\mathrm{0,-}}^{k,q} =&\frac{e^{i q y-i k(\epsilon)x}}{\sqrt{\cos\alpha(\epsilon)}}
\left(\begin{array}{c}
	e^{i \alpha(\epsilon)/2} \\
	-e^{-i \alpha(\epsilon)/2}
	\end{array}\right) \,.
\end{align}
Here the dispersion is given by $\epsilon=\hbar v_F \sqrt{q^2+k^2}$.
The wave vector $k(\epsilon)$ and the angle $\alpha(\epsilon)$
are defined as
\begin{align}
  \alpha(\epsilon)=&\arcsin\left(\frac{\hbar v_F q}{\epsilon +eV_g} \right)\\
  k(\epsilon)=&\frac{\epsilon +eV_g}{\hbar v_F } \cos\left(\alpha(\epsilon)\right)\,.
  \label{angel}
\end{align}
Therewith, and neglecting transverse momentum due to high doping, we have the basis states
\begin{align}
{{\Psi}}_{\mathrm{0,+}}^{k,0}&=\frac{e^{i k(\epsilon)x}}{\sqrt{2}}
\left(\begin{array}{c}
	1 \\
	1
	\end{array}\right)\\
{{\Psi}}_{\mathrm{0,-}}^{k,0}&=\frac{e^{-i k(\epsilon)x}}{\sqrt{2}}
\left(\begin{array}{c}
	1\\
	-1
	\end{array}\right) 
\end{align}
in the leads. Additionally shifting the Fermi surface of the graphene sheet away
from the Dirac point, and thus changing the concentration of carriers,
is incorporated into the gate voltage $eV_g$. 
For $\left|\epsilon+eV_g\right| < \left|\hbar v_F q\right|$ 
we have evanescent modes,~\cite{KatsnelsonGuinea07} with imaginary $\alpha(\epsilon)$ and $k(\epsilon)$. 
Otherwise we have propagating modes and scattering is only at $x=0,L$.\\
Irradiating the two-terminal structure with a laser~\cite{Erbe06} can be
described by a harmonic ac-bias voltage with driving strength
$\alpha=eV_{ac}/\hbar \omega$ as discovered in the pioneering paper by
Tien and Gordon~\cite{TienGordon62}. Their theory can be incorporated
into the scattering formalism~\cite{ButtikerPretre93,PedersenButtiker98} and we are
applying it here to the two-terminal graphene structure.
We take the two valleys and two pseudo-spin states of the carbon lattice 
into account in the pre-factor of the current operator of reservoir $\eta$, which reads
\begin{align}
&	\hat{I}_{\eta}(t)=\frac{2 e W}{\pi \hbar}  \sum\limits_{\gamma,\delta=L,R}^{} \sum\limits_{l,k=-\infty}^{\infty}   \int\limits_{-\infty}^{\infty} d\epsilon d\epsilon' \int\limits_{0}^{\infty}dq J_{l}\left(\alpha_{\gamma}\right) J_{k}\left(\alpha_{\delta}\right) \nonumber\\
	&\times \hat{a}_{\gamma}^{\dagger}(\epsilon - l \hbar \omega) 	A_{\gamma \delta}(\eta,\epsilon,\epsilon') \hat{a}_{\delta}(\epsilon' - k \hbar \omega)e^{i(\epsilon-\epsilon')t/\hbar} \,.
	\label{currentop}
\end{align}
Indices $\gamma,\delta$ run over reservoirs $L,R$. Summation over all modes of y-momentum 
is replaced by an integral since $W\gg L$. Scattering is contained within the current matrix 
${A}_{ \gamma \delta}(\eta, \epsilon, \epsilon')=\delta_{\eta \gamma} \delta_{\eta \delta} - s^*_{\eta \gamma}(\epsilon) s_{\eta \delta}(\epsilon')$ 
 of a current between leads $\gamma$ and $\delta$ measured in lead $\eta$ via the energy-dependent
scattering-matrix   
\begin{figure}[tb]
	\centering
		\includegraphics[width=0.72\columnwidth]{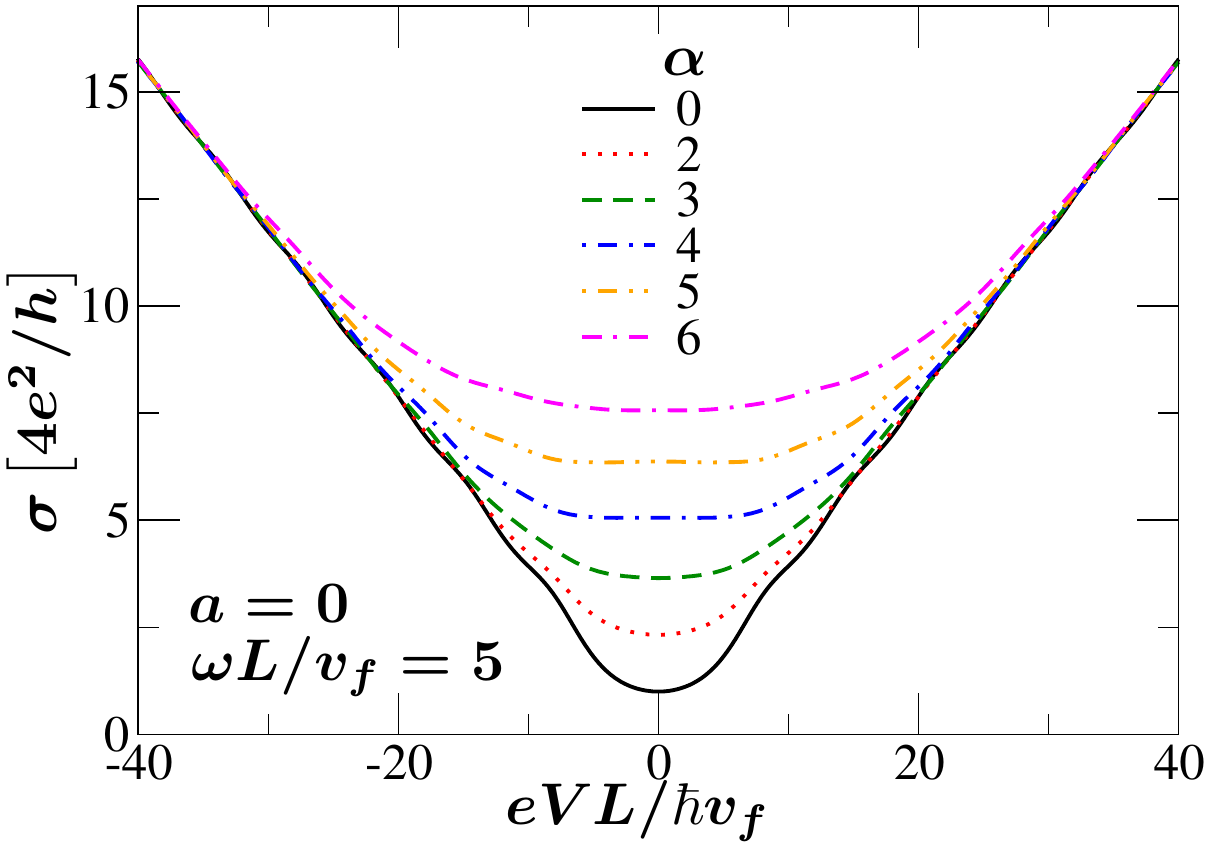}\\
		\includegraphics[width=0.72\columnwidth]{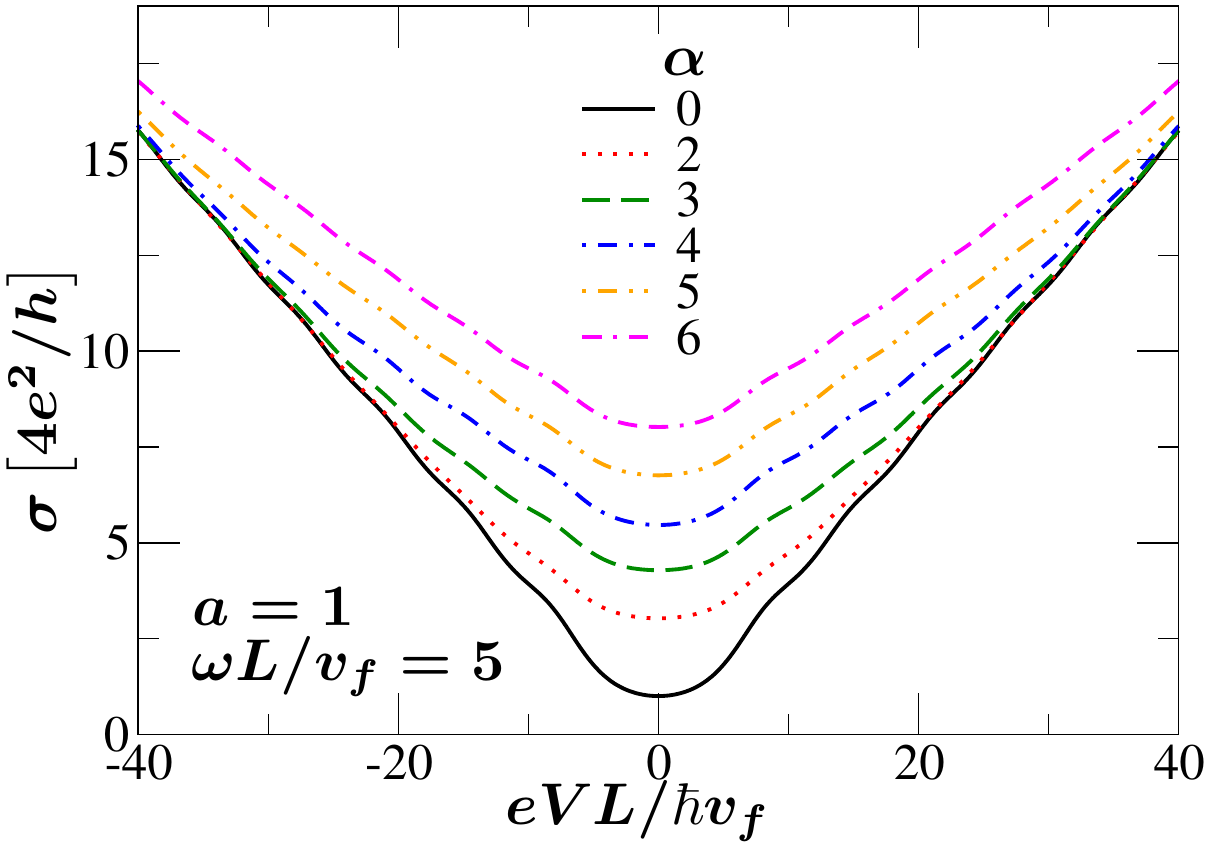}
		\caption{(color online) Left: Conductivity $\sigma(\omega,\alpha)=(L/W) G(\omega,\alpha)$ 
							as a function of 
							dc-voltage applied across the two-terminal setup. 
							We show curves for various 
							ac-driving strengths $\alpha$ applied to a) both reservoirs ($a=0$)
							and b) to the left reservoir only ($a=1$) with $\omega L/v_f=5$. 
		}
	\label{fig3}
\end{figure}
\begin{equation}
s(\epsilon)=\left(
	\begin{array}{cc}
	r(\epsilon) & t'(\epsilon) \\
	t(\epsilon) & r'(\epsilon)
	\end{array} 
	\right)	 \,.
\end{equation}
The scattering matrix connects in- and outgoing scattering states at the two barriers and
is calculated in Appendix \ref{app:bounds} by matching the wave functions at $x=0,L$. 
Here we write the results for reflection and transmission amplitudes in an alternative version:  
\begin{align}
t(\epsilon)&=\frac{2e^{ i k(\epsilon) L} \left(1+e^{2 i \alpha(\epsilon) }\right)}{e^{2 i k(\epsilon) L}\left(1-e^{i \alpha(\epsilon)} \right)^2+\left(1+ e^{i \alpha(\epsilon)} \right)^2}\\
r(\epsilon)&= \frac{\left(e^{2 i k(\epsilon) L}-1\right) \left(e^{2 i \alpha(\epsilon) }-1\right)}{e^{2 i k(\epsilon) L} \left(1-e^{i \alpha(\epsilon) }\right)^2+\left(1+e^{ i \alpha(\epsilon) }\right)^2}\,.
\label{transmission_exp}
\end{align}
We assume identical scattering for quasi-particles incident from left and right, 
so  $t(\epsilon)=t'(\epsilon)$ and $r(\epsilon)=r'(\epsilon)$.
$r(\epsilon)$ vanishes if $k(\epsilon)= \pi n/L$, with integer $n$. 
The corresponding modes in $y$-direction are determined by 
\begin{equation}
	q=\left[\left(\frac{\epsilon }{\hbar v_F}\right)^2-\left(\frac{\pi n}{L} \right)^2\right]^{1/2}\, ,
	\label{qjumps}
\end{equation} 
giving rise to special features of the current fluctuations, going along with the phase 
jumps of $\pi L/\hbar v_F$ in $r(\epsilon)$ we discuss later on.
At the Dirac point transmitted quasi-particles at perpendicular 
incidence perform Klein tunneling via evanescent modes,
leading to finite transmission probability $T(\epsilon)=t^{\dagger}(\epsilon)t(\epsilon)$ 
at small transverse momentum, see Fig.~\ref{fig2}. 
\section{Differential conductance \label{sec:conductance}} 
\begin{figure}[tb]
	\centering
	\includegraphics[width=0.97\columnwidth]{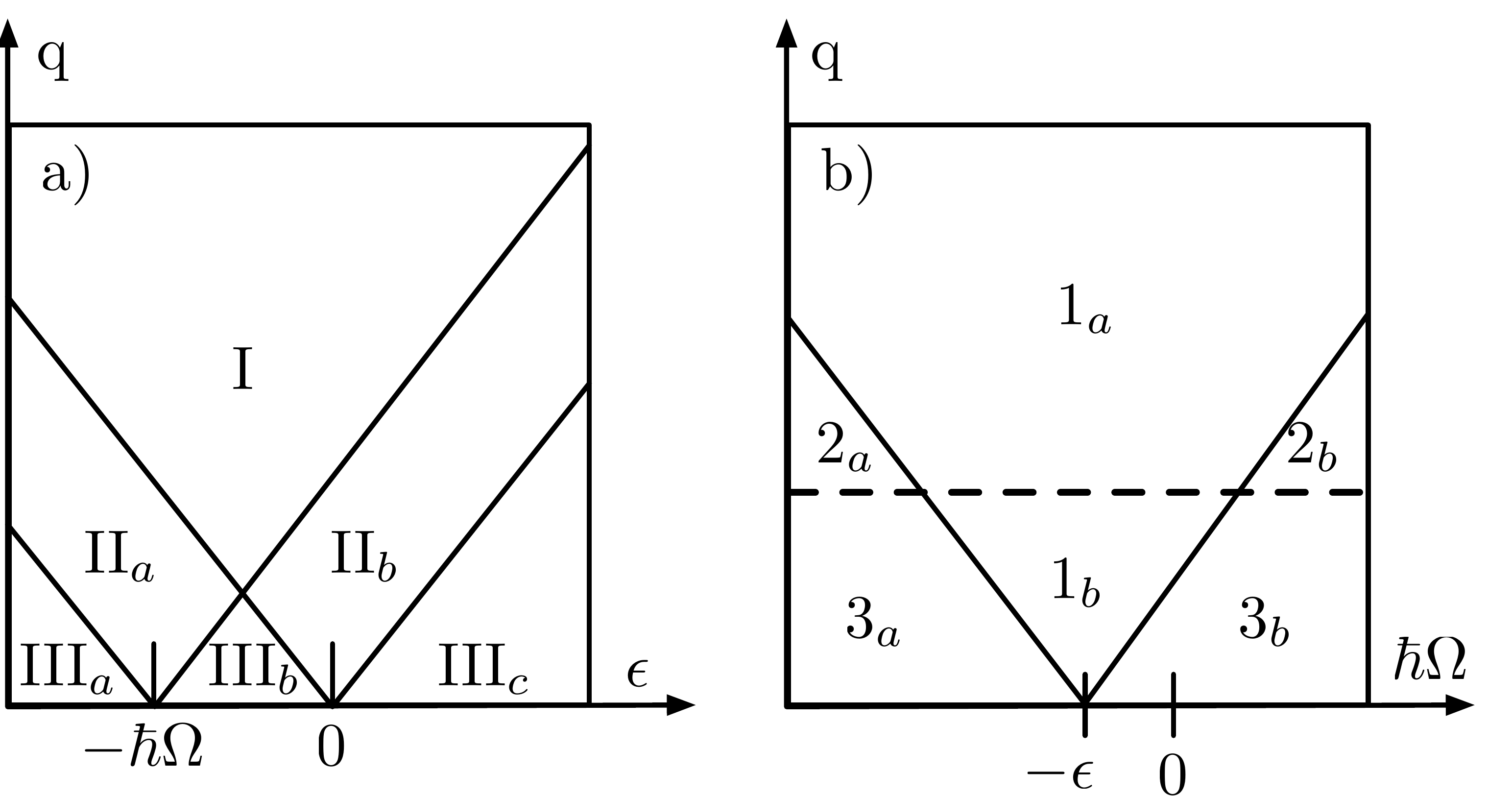}\\
	\caption{(color online) Schematic view of the different regions which occur in 
							the integrands of the correlators contributing to the 
							finite-frequency shot noise spectrum.}
	\label{fig4}
\end{figure}
Since the average current has only a zero-frequency component, 
PAT events in the conductance\cite{Snyman08,Peres06} can only be studied 
by inducing photon-exchange via a time-dependent voltage as it is, 
for example, generated by irradiating the setup with a laser beam. 
Different polarizations of the coupled light field lead to different ac-driving in left and right leads. 
Such an asymmetry can be described by a parameter $a \in \left[-1,1\right]$ which varies the 
driving in the leads via $\alpha_{L/R}=\frac{a \pm 1}{2}\alpha \equiv \frac{V_{\mathrm{ac,L/R}}}{\hbar \omega}$.
We call the driving symmetric (in the amplitudes $V_{\mathrm{ac,L/R}}$) if $a=0$ and asymmetric if $a= \pm 1$. 
For clearness we will only discuss  $a=0,\pm 1$ since intermediate values are just a 
mixture of those limiting cases.
For arbitrary $a$ the  differential conductance can be derived from 
Eqn.(\ref{currentop}) by taking the statistical- and time average and differentiating 
with respect to voltage. At $k_BT=0$ it reads
\begin{align}
	   G(\omega,\alpha)  =&\frac{2e^2 W}{\hbar}  \int\limits_{0}^{\infty}dq \sum\limits_{m=-\infty}^{\infty} \left( J_{m}^2\left(\alpha_L\right) \left|t (m\hbar \omega+\frac{eV}{2})\right|^2 \right. \nonumber\\
	&\left.+ J_{m}^2\left(\alpha_R\right)\left|t (m\hbar \omega-\frac{eV}{2})\right|^2 \right) \,.
\end{align}
Different orders $m$ of PAT do not mix but have to be summed up resulting 
in independent contributions $G_m(\omega,\alpha)$ to differential conductance.
Since $G(\omega,\alpha)$ only depends on the Besselfunctions squared, 
these pre-factors will always be positive. The influence of the driving 
strength $\alpha$ on conductivity $\sigma(\omega,\alpha)= (L/W) G(\omega,\alpha)$  
as a function of dc-bias is plotted in Fig.~\ref{fig3}.
PAT events lead to a substantial enhancement of the conductivity around zero dc-bias, 
because more channels are available in comparison to the case without time-dependent voltages. 
At large dc-bias voltages this effect gets negligible since the transmission probability 
of the graphene sheet, see Fig.~\ref{fig2}, is not vanishing at large energies. 
Thus, those contributions built a dominant background. 
Conductance at arbitrary dc- and ac-bias is a sum of two integrated  
transmission probabilities, where the integrand exhibits crossings of the two independent interference patterns,
as in region III$b$ in Fig.~\ref{fig4} a).   
Each $G_m(\hbar \omega) $ shows a transition from a region with an oscillating, but in average not increasing
contribution to conductance for dc-bias voltages $|eV/2| < |m\hbar \omega|$, 
to a regime with a linear increasing background at larger dc-bias voltages.
The photon-energy $m\hbar \omega$ introduces a phase shift in the oscillations of $G_m(\omega,\alpha)$
as a function of dc-bias voltage,
so for different $m$ we can have local minima or maxima at $eV=0$. 
After summation, conductivity can also show a local minimum or maximum at $eV=0$,  
as it can be observed for the various values of $\alpha$ in Fig.~\ref{fig3} a). 
If $|a|$ tends to one this effect is hidden behind the contribution from the terminal where 
driving gets small, as in Fig.~\ref{fig3} b) with $a=1$.
From the oscillations with period proportional to $L$, we expect no measurable effect on conductivity 
or shot-noise~\cite{DiCarlo08,Danneau08}, as in the case without ac-driving and for the zero-frequency Fano factor. 
In the scattering approach they are simply because the transmission function oscillates as a function of energy. 
But imperfections of real samples, as impurities~\cite{Titov07} or lattice-mismatch, lead to scattering events. 
Due to this randomizing effect on the path-lengths for propagating quasi-particles the 
calculated oscillations are averaged out in experiment~\cite{DiCarlo08,Danneau08}.
\section{Frequency-dependent shot noise \label{sec:shotnoise}}
 To get full informations on current-current correlations we study the
 non-symmetrized noise-spectrum as it can be detected by an
 appropriate measurement device in the quantum regime.~\cite{Schoelkopf97,Lesovik97,GavishImry00,GavishImry02,Beenakker01,Reydellet03,Nazarov08,Gabelli08,Gabelli09,Reulet03,Zakka-Bajjani09,EngelLoss04,EntinWohlman07,Rothstein08,Schonenberger03,Aguado00,Brandes04}\\

We allow harmonic ac-driving $eV_{ac} \cos(\omega t)$ in the leads,  
so in Fourier space the current-current correlations are defined as
\begin{equation}
  S_{\alpha \beta}(\Omega, \Omega', \omega)= \int\limits_{-\infty}^{\infty} dt dt' 
  S_{\alpha \beta}(t,t',\omega)
  e^{i \Omega t + i \Omega' t'} \, .   
  \label{noisedef}
\end{equation}
The non-symmetrized shot noise correlates currents at two times:
\begin{equation}
S_{\alpha \beta}(t,t',\omega)=\left\langle \Delta \hat{I}_{\alpha}(t) \Delta \hat{I}_{\beta}(t') \right\rangle
\end{equation}
with variance $\Delta \hat{I}_{\alpha}(t) = \hat{I}_{\alpha}(t) -
\langle\hat{I}_{\alpha}(t)\rangle$. 
Of experimental interest are the fluctuations on timescales large
compared to the one defined by the driving frequency $\omega$.  Thus,
as in~\cite{PedersenButtiker98}, we introduce Wigner coordinates
$t=T+\tau/2$ and $t'=T-\tau/2$ and average over a driving period
$2\pi/\omega$.  Then, the noise spectrum is defined by the quantum
statistical expectation value of the Fourier-transformed
current-operator $\hat{I}_{\alpha}(\Omega)$ via $S_{\alpha
  \beta}(\Omega, \Omega',\omega)=2\pi
S_{\alpha
  \beta}(\Omega,\omega)\delta(\Omega+\Omega') =
\langle\hat{I}_{\alpha}(\Omega)\hat{I}_{\beta}(\Omega')\rangle$.
$S_{\alpha \beta}(\Omega,\omega)$  is nothing but the Fourier transform of $S_{\alpha \beta}(\tau,\omega)$.
Similarly, in the case without ac-driving the noise is only a function
of relative times $\tau=t-t'$. In order to keep notation short,
in the dc-limit we write $S_{\alpha \beta}(\Omega):=S_{\alpha
  \beta}(\Omega,\omega=0)$.
To get a deeper insight into the underlying processes of charge-transfer 
we split the noise into four possible correlators~\cite{GavishImry02}, defined by
\begin{equation}
	S_{LL}(\Omega,\omega):=\underset{\alpha, \beta =L,R}\sum   C_{\alpha \rightarrow \beta}(\Omega,\omega) \,.
\end{equation}
The correlators itself can be seen as the building-blocks of noise spectra where 
different combinations describe noise detected by corresponding measurement setups.~\cite{GavishImry00,GavishImry02} 
First we discuss  $S_{LL}(\Omega):=S_{LL}(\Omega,\omega=0)$, the case when no ac-driving is present.
We also skip $\omega$ in the arguments of the correlators. 
Then evaluation of Eqn. (\ref{noisedef}) 
at $k_BT=0$ leads to the expressions:
\begin{subequations}
	\label{correlators}
\begin{align}
	C_{L \rightarrow L}(\Omega)=& 
	\frac{ e^2  \Theta(\hbar \Omega)}{2 \pi \hbar} \underset{\mu_L-\hbar \Omega}{\overset{\mu_L}\int}  d\epsilon \underset{-\infty}{\overset{\infty}\int}  dq \,
	 \left|r^*(\epsilon)r(\epsilon+\hbar\Omega) - 1 \right|^2  \label{correlators:LtoL}\\
	C_{R \rightarrow R}(\Omega)=&
	\frac{e^2  \Theta(\hbar \Omega)}{2 \pi \hbar} \underset{\mu_R-\hbar \Omega}{\overset{\mu_R}\int}  d\epsilon \underset{-\infty}{\overset{\infty}\int}  dq \, 
 	T(\epsilon)T(\epsilon+\hbar\Omega)\label{correlators:RtoR}\\
	C_{L \rightarrow R}(\Omega)=&
	 \frac{e^2 \Theta(\hbar \Omega-eV)}{2 \pi \hbar} \underset{\mu_R-\hbar \Omega}{\overset{\mu_L}\int}  d\epsilon\underset{-\infty}{\overset{\infty}\int}  dq \, 
 	R(\epsilon)T(\epsilon+\hbar\Omega)\label{correlatorsL:toR}\\
	C_{R \rightarrow L}(\Omega)=& 
	\frac{e^2 \Theta(\hbar \Omega+eV)}{2 \pi \hbar} \underset{\mu_L-\hbar \Omega}{\overset{\mu_R}\int}  d\epsilon \underset{-\infty}{\overset{\infty}\int}  dq \,
	T(\epsilon)R(\epsilon+\hbar\Omega)\, . \label{correlators:RtoL}
\end{align}
\end{subequations}
At finite dc-bias voltages correlations with initial and final state related to the 
measurement terminal $L$ are special in the sense that they can not be written in terms of
probabilities at finite frequency. For symmetrized noise, B\"uttiker~\cite{Buttiker92} discussed the essential 
role of the complex reflection amplitudes in elastic electron transport and how they determine the equilibrium current fluctuations.
In the quantum regime at $k_BT=0$, the equilibrium fluctuations are given by
\begin{align}
	S_{LL}(\Omega)=&	\frac{e^2}{2 \pi \hbar} \Theta(\hbar \Omega) \underset{-\hbar \Omega}{\overset{0}\int}  d\epsilon\underset{-\infty}{\overset{\infty}\int}  dq \nonumber\\
&	\left(2-	r^*(\epsilon)r(\epsilon+\hbar\Omega)-r^*(\epsilon+\hbar\Omega)r(\epsilon)  \right) \,.
\label{equilibrium}
\end{align}
For finite dc-bias the reflection amplitudes entering $C_{L\rightarrow L}(\Omega)$
play the same essential role as in equilibrium, 
in the sense that finite-frequency current fluctuations are non-zero even for vanishing transmission.
The combination of scattering-matrices of the correlators integrands which enter in the current-current cross-correlation spectrum
\begin{equation}
	S_{LR}(\Omega,\omega):=\underset{\alpha, \beta =L,R}\sum   C^{\mathrm{c}}_{\alpha \rightarrow \beta}(\Omega,\omega) 
\end{equation}
are substantially  different than in the ones for the auto-terminal noise. Most of all, at finite frequency none of the 
complex correlators can
be written as an integral over transmission- or reflection probabilities: 
\begin{subequations}
	\label{correlatorscross}
\begin{align}
		C^{\mathrm{c}}_{L \rightarrow L}(\Omega)=&  \frac{e^2 \Theta(\hbar \Omega)}{2 \pi \hbar} \underset{\mu_L-\hbar \Omega}{\overset{\mu_L}\int} d\epsilon\underset{-\infty}{\overset{\infty}\int}  dq \, \nonumber \\
&	t^*(\epsilon+\hbar\Omega) 	t(\epsilon) \left[1- r^*(\epsilon)r(\epsilon+\hbar\Omega) \right]\label{correlatorscross:LtoL} \\
		C^{\mathrm{c}}_{R \rightarrow R}(\Omega)=&  \frac{e^2 \Theta(\hbar \Omega) }{2 \pi \hbar} \underset{\mu_R-\hbar \Omega}{\overset{\mu_R}\int} d\epsilon \underset{-\infty}{\overset{\infty}\int}  dq \, \nonumber \\
&
	 		t^*(\epsilon) t(\epsilon+\hbar\Omega) \left[1- r^*(\epsilon+\hbar\Omega)r(\epsilon) \right]\label{correlatorscross:RtoR}\\
		C^{\mathrm{c}}_{L \rightarrow R}(\Omega)=& \frac{-e^2 \Theta(\hbar \Omega-eV)}{2 \pi \hbar} \underset{\mu_R-\hbar \Omega}{\overset{\mu_L}\int} d\epsilon \underset{-\infty}{\overset{\infty}\int}  dq \,\nonumber \\
&
	 	r^*(\epsilon)t(\epsilon)  r^*(\epsilon+\hbar\Omega)t(\epsilon+\hbar\Omega)\label{correlatorscross:LtoR}\\
		C^{\mathrm{c}}_{R \rightarrow L}(\Omega)=&  \frac{-e^2 \Theta(\hbar \Omega+eV)}{2 \pi \hbar}  \underset{\mu_L-\hbar \Omega}{\overset{\mu_R}\int} d\epsilon \underset{-\infty}{\overset{\infty}\int}  dq \, \nonumber \\
&
		t^*(\epsilon)r(\epsilon)t^*(\epsilon+\hbar\Omega)r(\epsilon+\hbar\Omega)\label{correlatorscross:RtoL}
\end{align}
\end{subequations}
Unlike for symmetrized noise, quantum noise~\cite{Nazarov08,Schonenberger03} spectra discriminate between
photon absorption ($\Omega > 0$) and emission ($\Omega < 0$) processes between
quasi-particles in graphene and a coupled electric field~\cite{GavishImry00,GavishImry02,Aguado00,Brandes04,Marcos11}.
Energy for photon emission has to be provided by the voltage source, so at $k_BT=0$ the Heaviside-Theta functions ensure that only 
terms satisfying this condition contribute at negative frequencies. In the dc-limit, our choice of chemical potentials 
$-\mu_L=\mu_R=eV/2 > 0$ and the fact that the measurement is performed at reservoir $L$, leaves only 
$C^{\mathrm{c}}_{R \rightarrow L}(\Omega) \ne 0$ if $\Omega  \le 0$. When additional ac-voltages are present none
of the correlators of Eqn.(\ref{thenoise}) is given in terms of probabilities and integration boundaries are
changed by $\pm m\hbar \omega$. Then all correlators can contribute at frequencies $\Omega<0$. 
\begin{figure}[tbp]
  \centering
  \includegraphics[width=0.95\columnwidth]{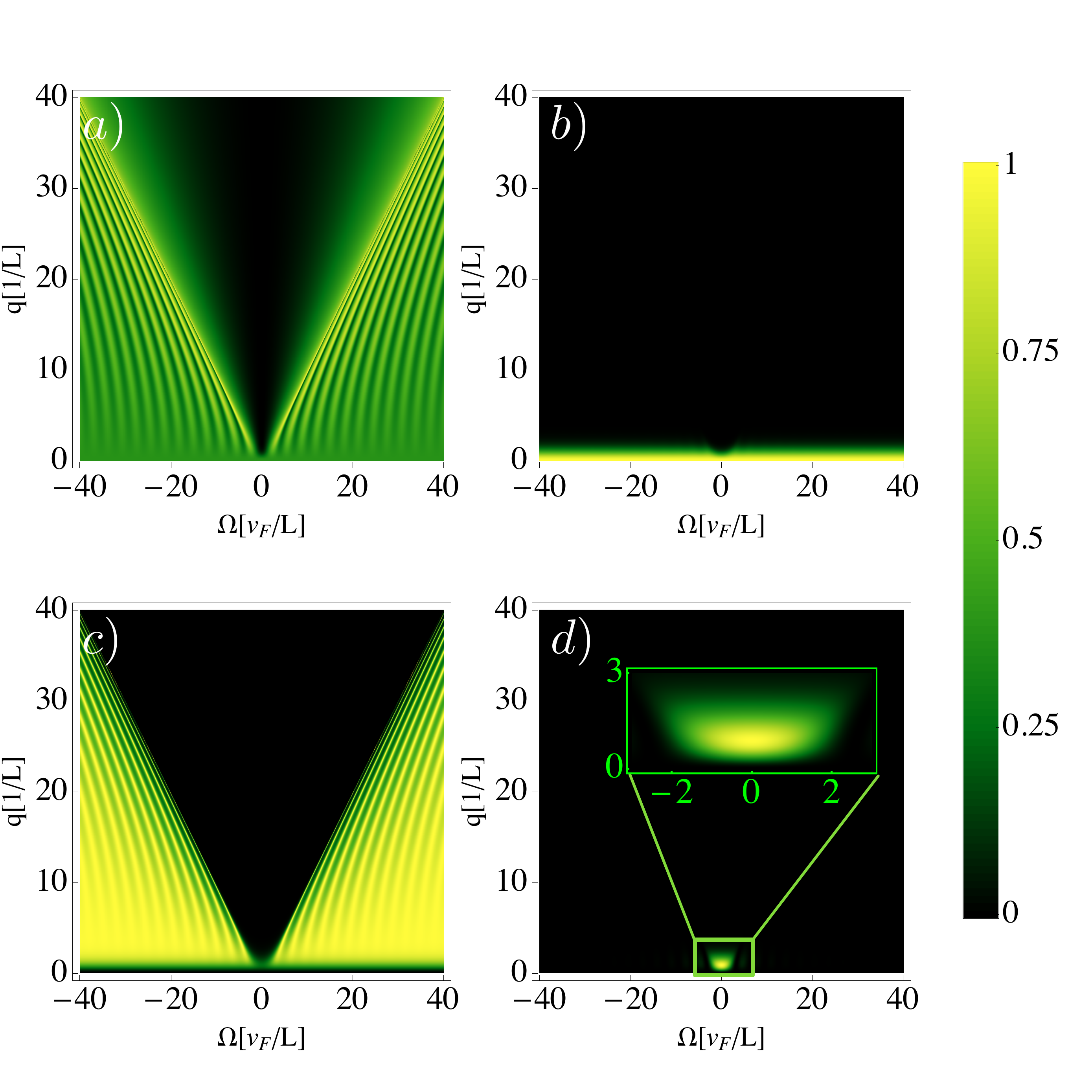}
  \caption{(color online) Real parts of integrands of the four correlators
	(Eqn.~\ref{correlators})
    contributing to the shot-noise, namely a) $0.25|1-r^*(\epsilon)r(\epsilon+\hbar \Omega)|^2$, b) $T(\epsilon)T(\epsilon+\hbar \Omega)$, c) $R(\epsilon)T(\epsilon+\hbar \Omega)$ and d) $R(\epsilon+\hbar \Omega)T(\epsilon)$. Here the energy is fixed
    $\epsilon = 0$ corresponding to vanishing dc-bias.  The correlator
    in a) cannot be written in terms of a probabilities, except in the
    zero frequency limit the integrand results in
    $T^2(\epsilon)$. Correlator b) contains one transmission
    probability at zero energy that is only non-zero at small $q$.  
	Since for small transversal momentum $R(\epsilon)$
    decays as $q^{-2}$ the correlator c) tends to zero in this regime and
    otherwise mimics the behavior of $T(\epsilon)$. Integrand d) is
    also restricted to low transverse momentum because 
    $T(\epsilon)=0$ otherwise.\\
	}
  \label{fig5}
\end{figure}
\begin{figure}[tbp]
		\centering
		\includegraphics[width=0.9\columnwidth]{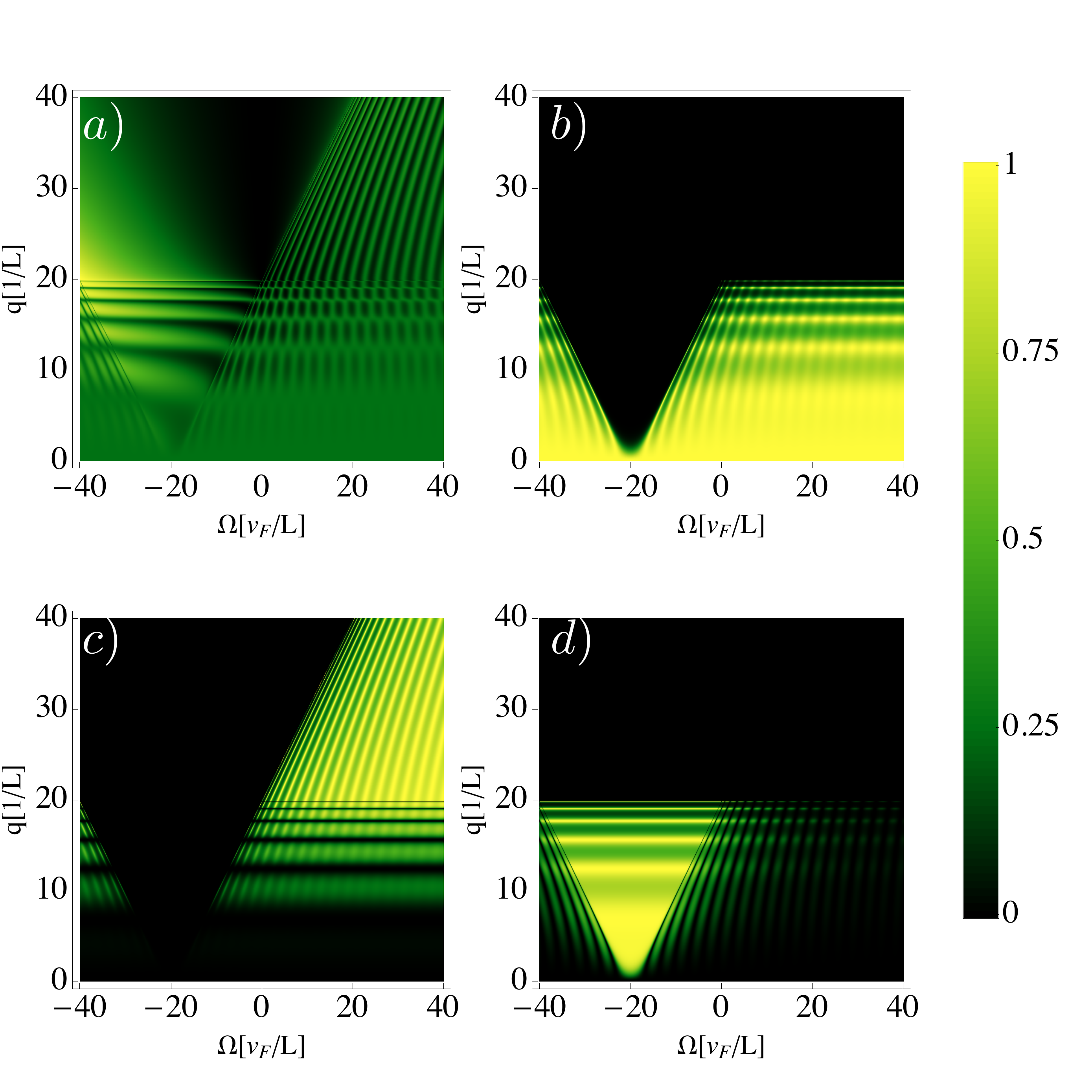}
		\caption{(color online) Real parts of integrands of  correlators 
			(Eqn.~\ref{correlators}, see also Fig.~\ref{fig5}) contributing to the shot-noise  
			for fixed energy $\epsilon L/\hbar v_F=20$. At finite $\epsilon$ there is an additional 
			interference pattern along $q$ if $\hbar v_F |q| < |\epsilon|$, 
			leading to phase jumps in the Integrand of correlator a), the one where initial and final 
			state belong to the measurement terminal $L$. When the integrand can be written as a product 
			of probabilities, see b)-d), the phase jumps are absent but 
			two independent interference patterns are found. 	}
	\label{fig6}
\end{figure}
\begin{figure}[tbp]
		\centering
		\includegraphics[width=0.9\columnwidth]{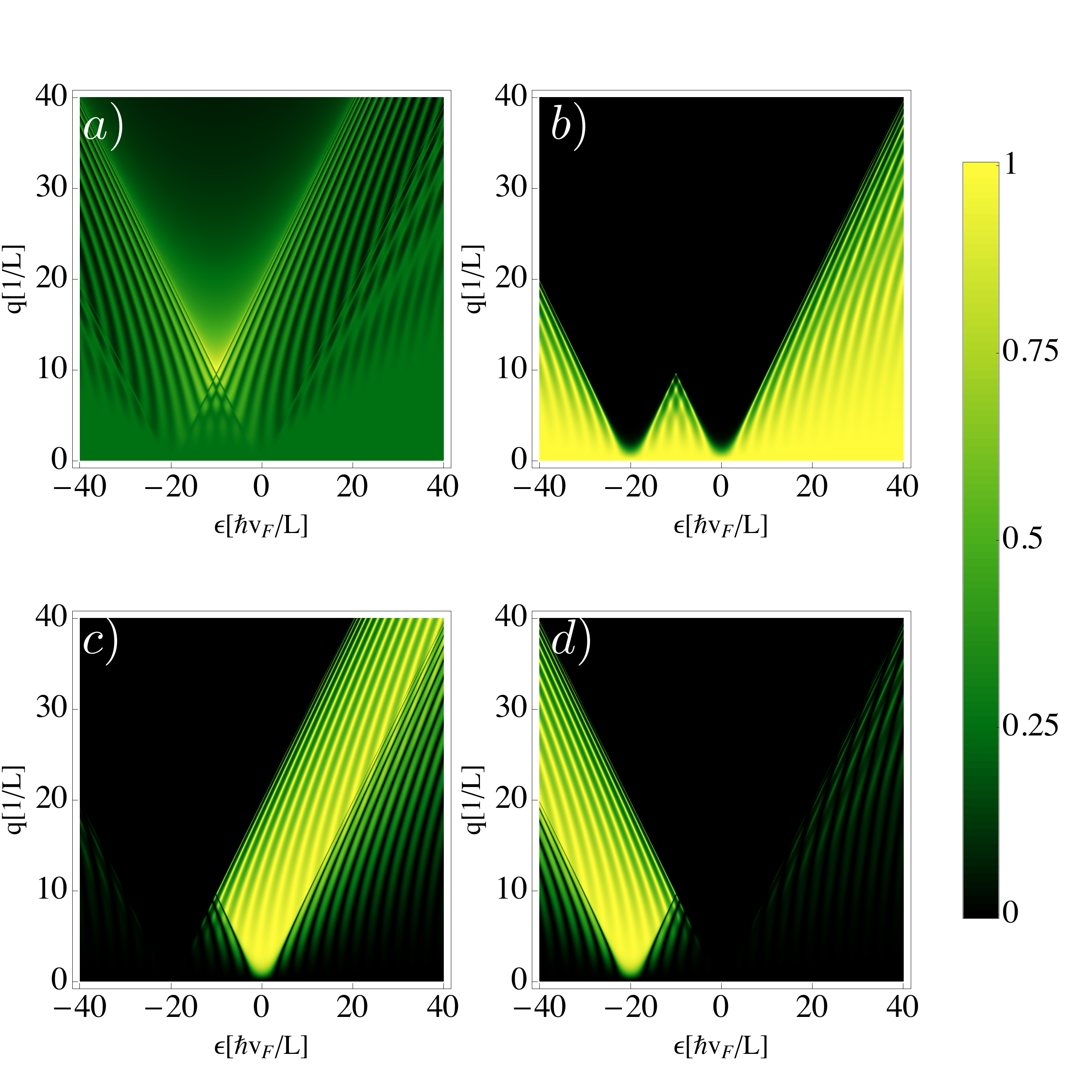}
		\caption{(color online) Real parts of integrands of the correlators (Eqn.~(\ref{correlators}), see also Fig.~\ref{fig5})
			contributing to the shot-noise  with fixed frequency $\Omega L/v_F=20$. 
			Analogous to Fig.\ref{fig6} but as a function of $(q,\epsilon)$.
			Phase jumps occur in the intervall $-\hbar \Omega<\epsilon<0$ in integrand a), 
			region III$_b$ of Fig.\ref{fig4}a).
			The interplay of the two interference patterns can also be observed at larger energies and 
			transverse momenta for $\hbar v_F |q|< |\epsilon|$, $\hbar v_F |q|< |\epsilon+ \hbar \Omega|$ in all integrands a)-d).  
		}
	\label{fig7}
\end{figure}
\begin{figure}[tbp]
		\centering
		\includegraphics[width=0.9\columnwidth]{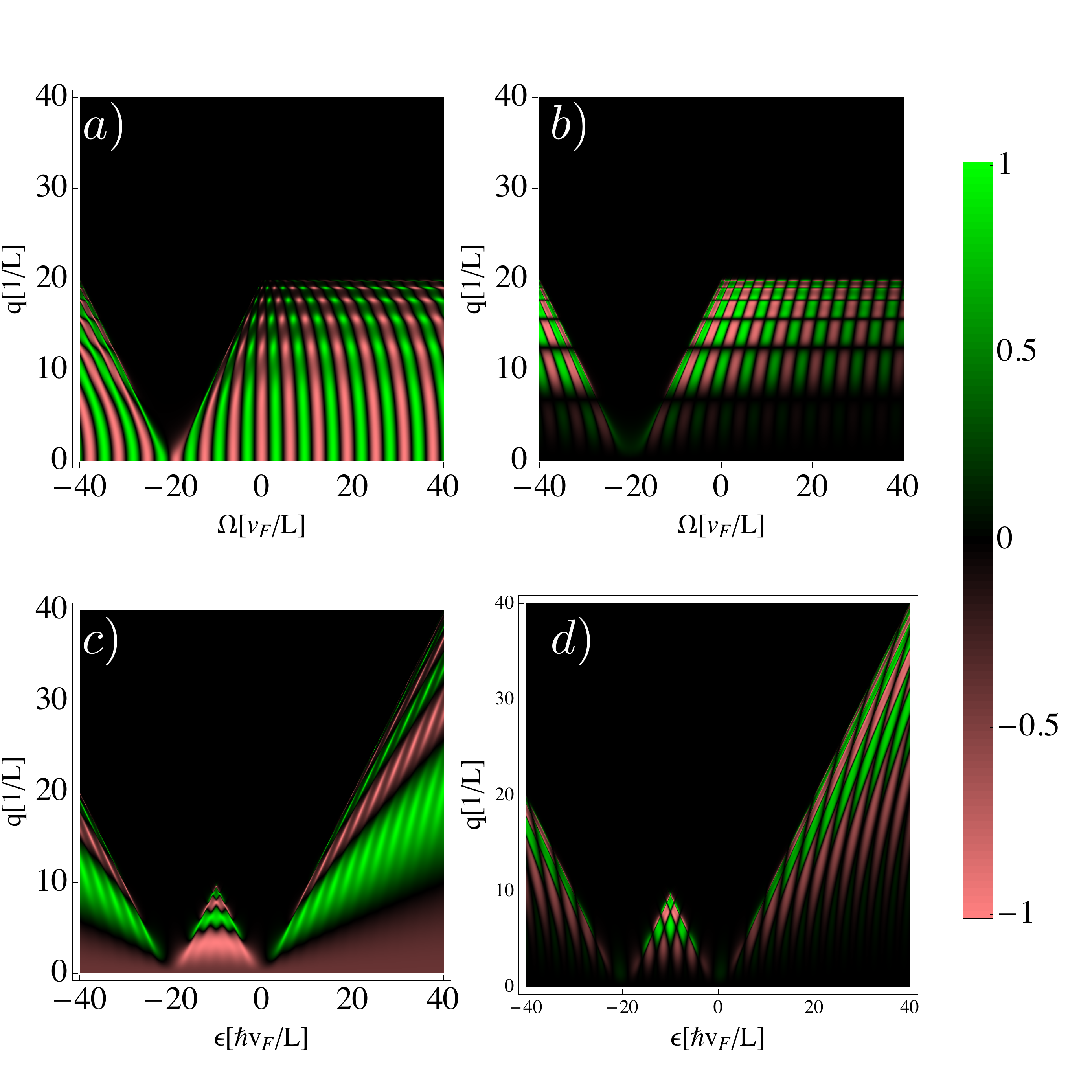}
		\caption{(color online) Real parts of integrands which appear in Eqn.~(\ref{correlatorscross}) 
			contributing to the cross-correlation shot-noise 
	 		for fixed energy $\epsilon L/\hbar v_F=20$ (top) and fixed frequency $\Omega L/v_F=20$ (bottom), 
			namely a),c) $\Re[t^*(\epsilon+\hbar\Omega)t(\epsilon)(r^*(\epsilon+\hbar\Omega)r(\epsilon)-1)]$ and
			b),d) $4\Re[r^*(\epsilon)t(\epsilon+\hbar\Omega)r^*(\epsilon+\hbar\Omega)r(\epsilon)]$.
			Due to symmetry reasons the integrands are identical when interchanging index labels $L,R$. 
			As a function of frequency  integrand a) leads to strongly oscillating  
			contributions to the noise spectrum. These oscillations are reduced due to 
			the alternating behavior along $q$ in cross-terminal contributions b). 
			In c), d) the integrands are plotted as a function of $(q,\epsilon)$ where they 
			reveal a similar structural difference.
			}
	\label{fig8}
\end{figure}
\begin{figure}[tbp]
		\centering
		\includegraphics[width=0.9\columnwidth]{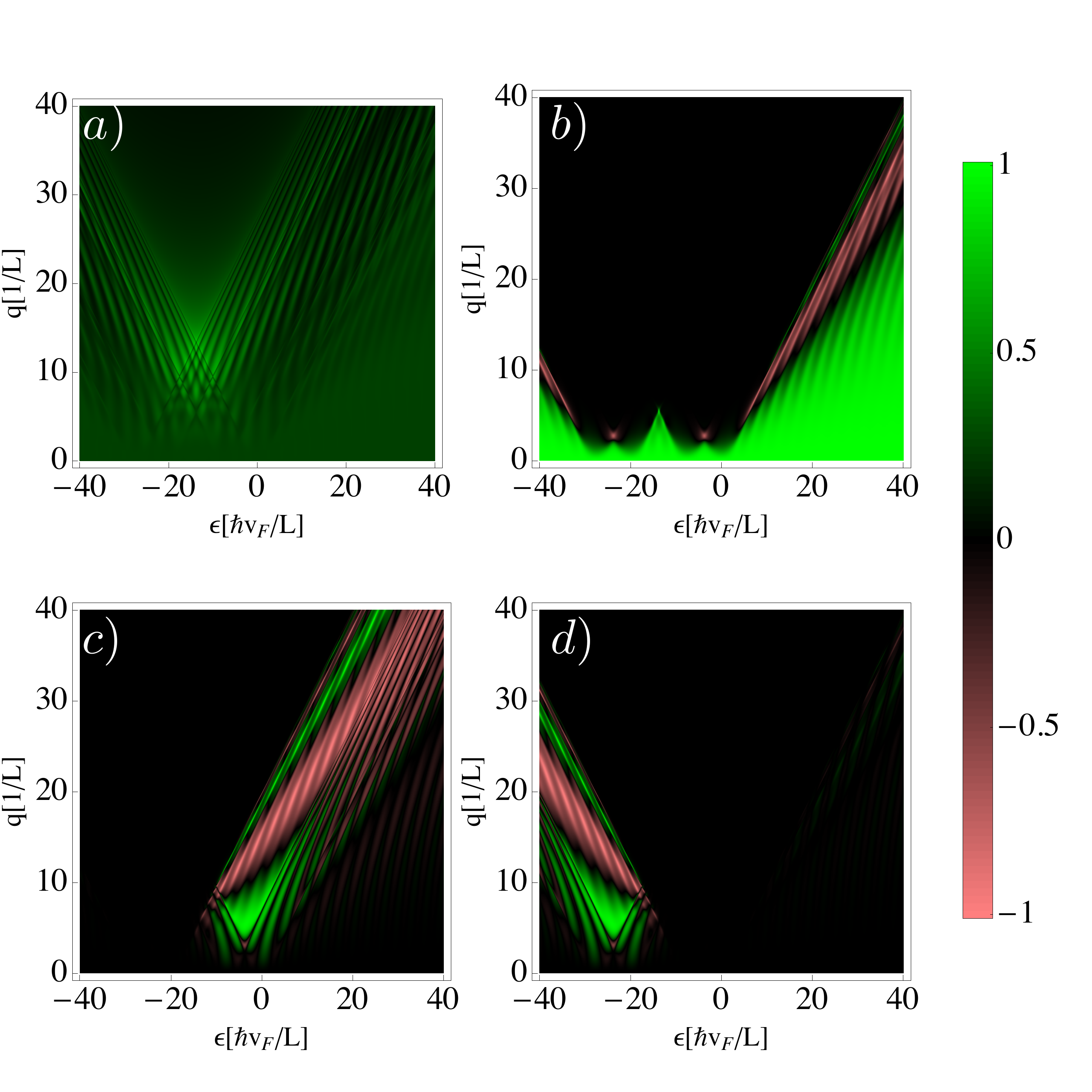}	
		\caption{	(color online) Real parts of integrands appearing in 
		    the correlators of Eqn.~(\ref{thenoise}) contributing to 
			the shot-noise for driving frequency $\omega L/v_F=7.5$ and fixed frequency $\Omega L/v_F=20$. 
			The integrands are a) $0.25(1-r^*(\epsilon)r(\epsilon+\hbar\Omega))(1-r^*(\epsilon+\hbar\Omega+\hbar\omega)r(\epsilon+\hbar\omega))$, 
			b) $t^*(\epsilon)t(\epsilon+\hbar\Omega)t^*(\epsilon+\hbar\Omega+\hbar\omega)t(\epsilon+\hbar\omega)$, 
			c) $r^*(\epsilon)t(\epsilon+\hbar\Omega)t^*(\epsilon+\hbar\Omega+\hbar\omega)r(\epsilon+\hbar\omega)$ and 
			d) $t^*(\epsilon)r(\epsilon+\hbar\Omega)r^*(\epsilon+\hbar\Omega+\hbar\omega)t(\epsilon+\hbar\omega)$.
			When two frequencies are present none of 
			the correlators can be written in terms of probabilities and additional 
			phase jumps come into play.
	}
	\label{fig9}
\end{figure}
\begin{figure}[tbp]
		\centering
		\includegraphics[width=0.9\columnwidth]{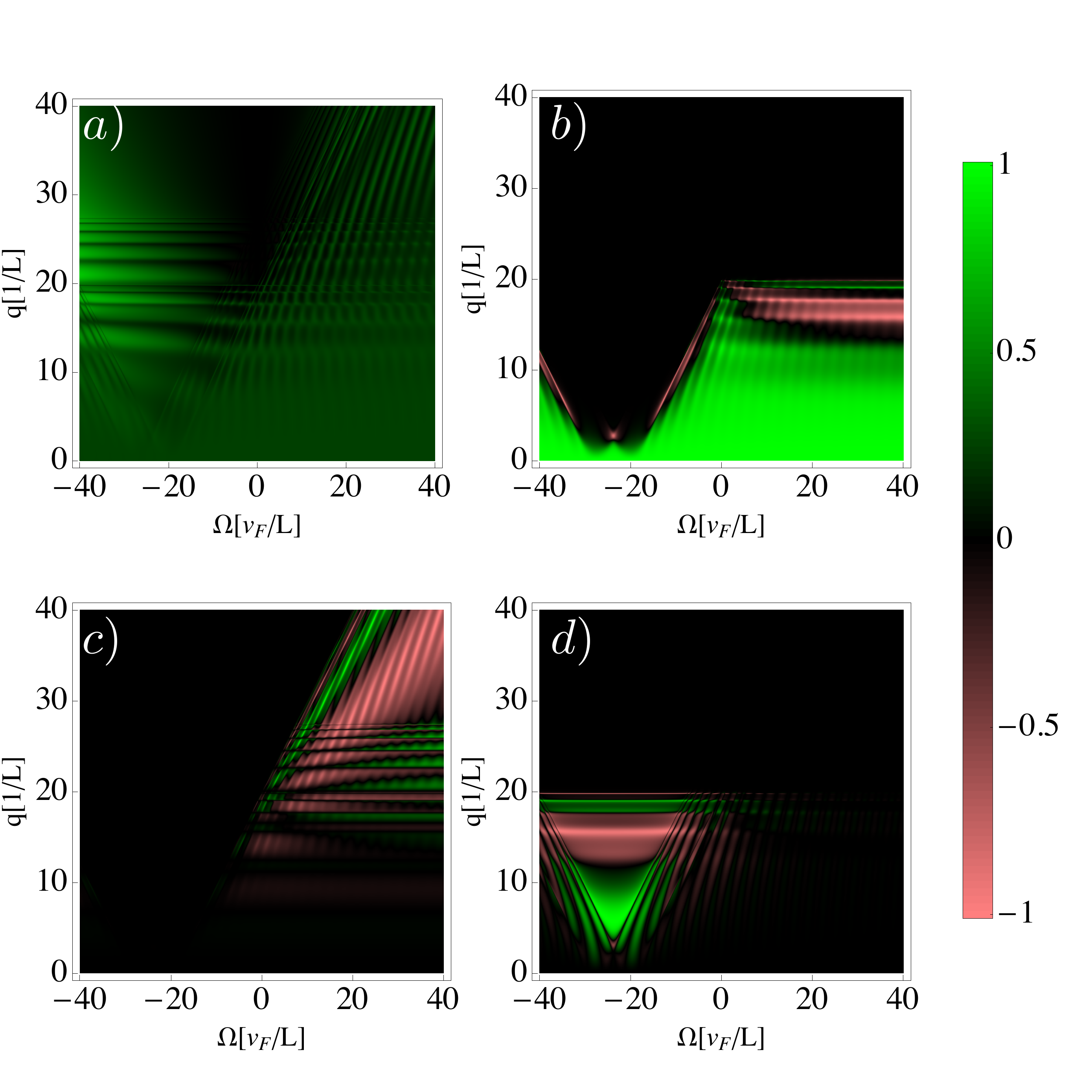}		
		\caption{ (color online) Real parts of integrands of the correlators contributing to the 
			shot-noise with driving frequency $\omega L/v_F=7.5$
			and	fixed energy $\epsilon L/\hbar v_F=20$. Analogous to Fig.~(\ref{fig9}), 
			but as a function of $(q,\Omega$). As a
			consequence of PAT horizontal interference lines occur for transverse 
			momenta $\hbar v_F|q|<|\epsilon+m\hbar \omega|$ as in figures a), c).}
	\label{fig10}
\end{figure}
\section{Qualitative discussion \label{sec:qualitative}}
A good starting point to interpret results for conductivity and  
shot-noise spectra is to examine the involved integrands in Eqs.~(\ref{correlators}) and (\ref{correlatorscross}). 
Figure~\ref{fig4} provides a schematic overview of the different 
regions occurring in the 2D-plots of Figs.~\ref{fig5}-\ref{fig10}.
We show the real parts of integrands either as a function of $(q,\epsilon)$ as in scheme ~\ref{fig4} a)  
or of $(q,\Omega)$  as in scheme~\ref{fig4} b). 
The former is divided by the four envelopes $q=|\epsilon|$ 
and $\hbar v_F q=|\epsilon+\hbar \Omega|$ into six areas: 
I, where the regimes II$_a$ and II$_b$ of evanescent modes are merging 
and the areas III$_a$,III$_a$, III$_a$ of propagating modes. 
Area III$_b$ is defined by the two lines with origins $(q=0,\epsilon=0$), ($q=0,\epsilon=-\hbar \Omega)$ 
and intersection $(\hbar v_Fq=\hbar \Omega/2,\epsilon=-\hbar \Omega/2)$.
Areas in scheme~\ref{fig4}b) are separated by $\hbar v_F q=|\epsilon + \hbar \Omega|$ 
and the dashed horizontal line $\hbar v_F q=|\epsilon|$. 
The transmission probability fits into this scheme when the horizontal separation is absent 
so we are left with areas 1$_a$ and 2$_{a/b}$. Then area 1$_a$
includes the black region of Fig.~\ref{fig2} where no transmission is possible, and the regime 
of evanescent modes with finite transmission probability for small $|\epsilon| < \hbar v_F |q|$ around $\epsilon=0$ due to Klein tunneling. 
In regimes 2$_{a/b}$ a hyperbolic shaped interference pattern with oscillations along $\epsilon$ is prominent, 
where  the period of oscillations is on the order of $\hbar v_f/L$ for small $\hbar v_F |q| \ll  |\epsilon|$. 
%
%
Figure~\ref{fig5} shows the relevant integrands of 
the four correlators $C_{\alpha \rightarrow \beta}(\Omega)$
contributing to the finite frequency quantum noise, plotted as a function of $(q,\Omega)$ when $\epsilon=0$. 
Then the imaginary part of $r^*(\epsilon)r(\epsilon+\hbar \Omega)$ leads to finite 
contributions in the region I$_a$ and I$_b$ in figure~\ref{fig5}a). 
$T(\epsilon=0)$ is only non-zero for small $q$, 
so integrands b) and d) vanish for large $q$. 
Since $R(\epsilon)=1-T(\epsilon)$, integrand c) vanishes when $q\rightarrow 0$ 
and otherwise resembles the shape of $T(\epsilon)$.\\
Finite $\epsilon$, as in Fig.~\ref{fig6}, introduces another 
interference pattern for propagating modes.
In region 1$_a$ non-zero values are possible and in 2$_a$ and 2$_b$ the usual interferences occur. 
For $q$-values below $\hbar v_F |q|=|\epsilon|$ this additional pattern can be seen in region 1$_b$. 
The interplay of both patterns leads to phase jumps of $\pi L/\hbar v_F$ in regions 3$_a$ and 3$_b$.
These phase jumps can be determined by requiring $|r^*(\epsilon) r(\epsilon+\hbar \Omega)-1|^2=1$ 
in Eq.(\ref{correlators:LtoL}), Fig.~\ref{fig6}a).
Therefor $r^*(\epsilon) r(\epsilon+\hbar \Omega)$ has to vanish, what is fulfilled by the transversal momenta
of Eq.~(\ref{qjumps}).
The condition $|r^*(\epsilon) r(\epsilon+\hbar \Omega)-1|^2=4$ for a 
maximum in the integrand leads to modes which experience Klein tunneling.  
Actually, this correlator can be written as integral over 
$1 + R(\epsilon)R(\epsilon+\hbar\Omega) -  2 [R(\epsilon)R(\epsilon+\hbar \Omega)]^{1/2} \cos(\Phi(\epsilon,\Omega))$ 
including a scattering-phase $\Phi(\epsilon,\Omega) = \text{Arg}\left[r^*(\epsilon)r(\epsilon+\Omega)\right]$. 
Thus it describes events containing the scattering-phase between time-reversed paths of 
electron-hole pairs separated by the photon energy $\hbar \Omega$ reflected back into the measurement terminal.
The effect of the phase shifts on the integrands interference patterns is also obvious 
in the $(q,\epsilon)$-plot of Fig.~\ref{fig7} a), region III$_b$. 
Figs.~\ref{fig7} b)-d) show a similar interference pattern
although the corresponding correlators are defined in terms of probabilities.\\
%
%
Concerning cross-correlation noise, the integrands occurring in Eqs.(\ref{correlatorscross}) show alternating patterns of 
positive and negative values. The ones which describe auto-terminal 
contributions  to $S_{LR}(\Omega)$ (Eqs.(\ref{correlatorscross:LtoL}) and (\ref{correlatorscross:RtoR})), 
as in Fig.~\ref{fig8}a), 
have an alternating  sign along $\Omega$. 
In the cross-terminal ones (Eqs.(\ref{correlatorscross:LtoR}) and (\ref{correlatorscross:RtoL})), 
as in Fig. ~\ref{fig8}b), the additional 
interference pattern along $q$ introduces another change of sign. 
Plots ~\ref{fig8} c) and d) show a similar behavior as functions of $(q,\epsilon)$.  
%
%
When ac-bias voltages introduce the driving frequency $\omega$, the integrands structures become even richer but also less clear, as in Fig.~\ref{fig9} and Fig.~\ref{fig10}. 
Then alternating signs in all contributions to auto-correlation noise are observed, except for the correlator with initial and final sates in the measurement terminal. 
This results in peculiar oscillatory features in the interference patterns at combinations of all 
involved energies $\epsilon,\hbar \Omega,m \hbar \omega$. Predicting the effect of such features on the noise 
spectra from the plotted integrands is then almost impossible because one still has to average over all possible 
energies and $q$-values by integration. 
\section{Auto-correlation noise \label{sec:autonoise}} 
In contrast to conductivity, the shot-noise spectrum in general couples 
different orders of PAT events, expressed by the product of 
four Besselfunctions of arbitrary order. But since the driving is fixed,
non-vanishing contributions exist only up to a certain order depending on the 
precise value of $\alpha$. When time-dependent voltages are present, current fluctuations of Eq.~(\ref{thenoise}) contain
products of four scattering matrices, each with a different energy argument. 
After performing the dc-bias limit only transitions between $\epsilon$ and $\epsilon + \hbar \Omega$ are left.
 \begin{figure}[tb]
	\centering
		\vspace{0.2cm}
		\includegraphics[width=0.97\columnwidth]{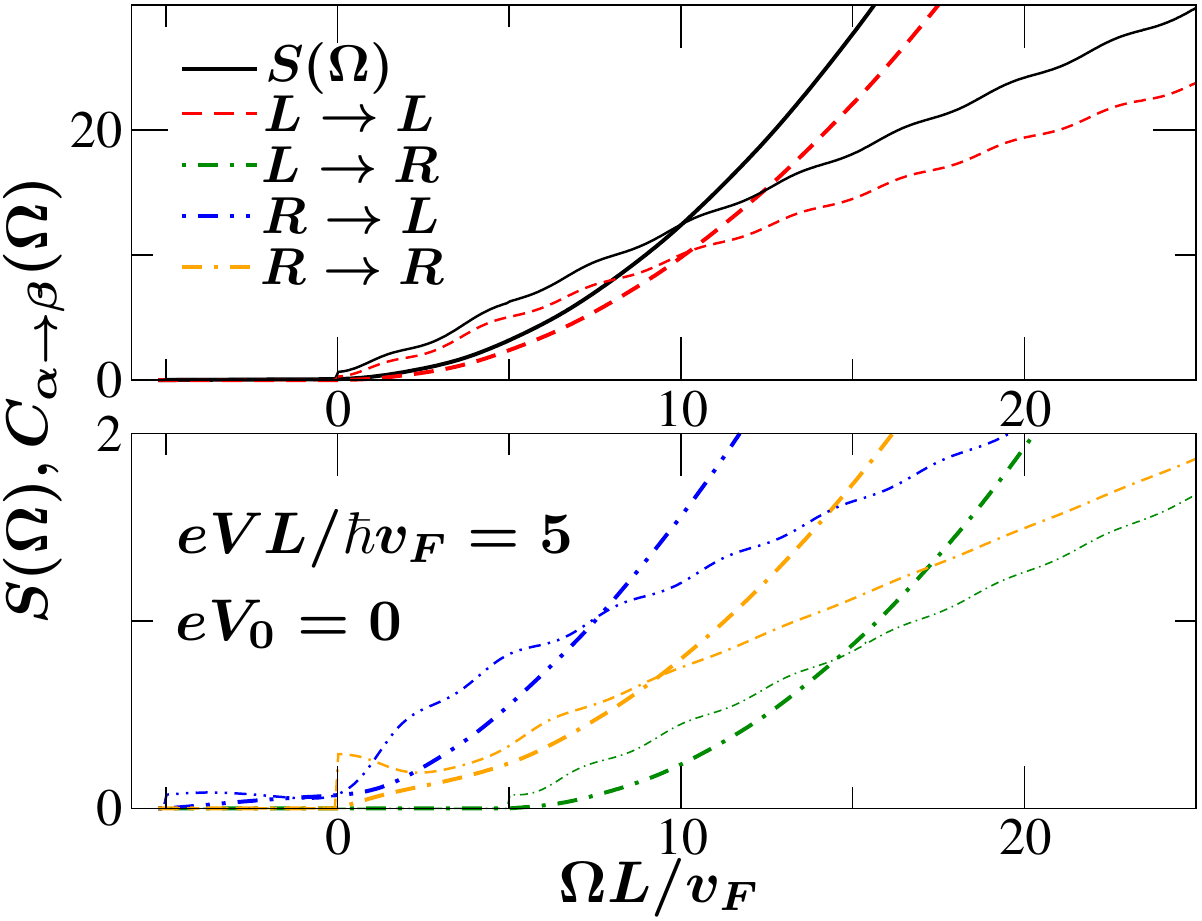}\\
		\vspace{0.6cm}
		\includegraphics[width=0.97\columnwidth]{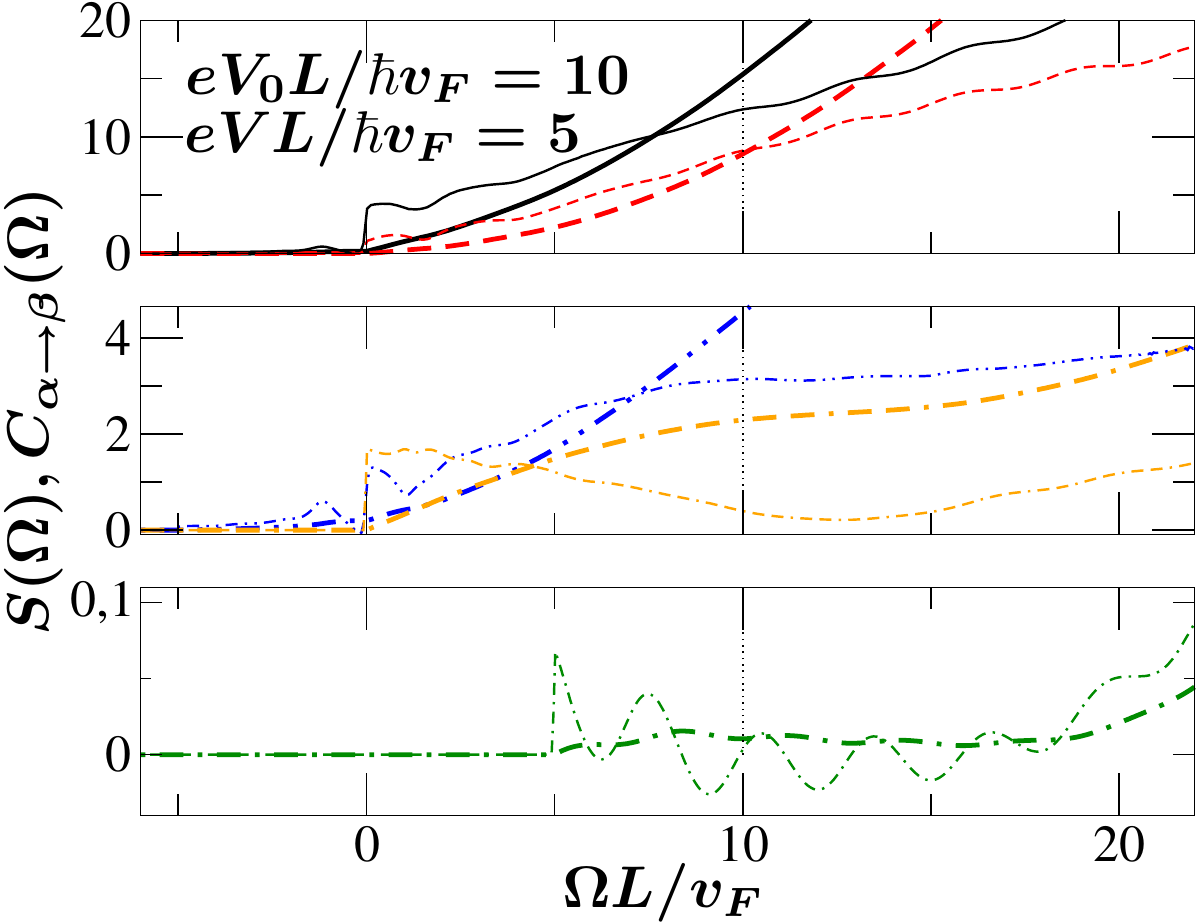}
		\caption{(color online) Real parts of auto-correlation noise spectrum in units of $2\pi\hbar/e^2$. 
			We compare a setup where dc-bias voltages are fixed symmetrically around 
			the Dirac point (top,) with the case when $eV_0L/\hbar v_F=2eV/\hbar v_F=10$ (bottom).
			Thick lines: Shot-noise and correlators. 
			Thin lines: Derivatives with respect to frequency. 
			Contributions from $C_{L\rightarrow L}(\Omega)$ 
			are dominant at positive frequencies. 
			Top: Special features in the derivatives are seen for frequencies $\hbar \Omega < eV$ 
			in the $R \rightarrow R$ contribution, when the lower bound of the energy-integration interval 
			approaches the Dirac point (compare to Figs.~\ref{fig6},~\ref{fig7}). 
			Bottom: The distance to the Dirac point is increased by the offset voltage. 
			Therefore oscillatory features appear in a  larger frequency interval and in all four correlators, 
			since integration boundaries in all contributions are crossing the 
			Dirac-point with increasing $\Omega$.}
	\label{fig11}
\end{figure}
 \begin{figure}[tb]
	\centering
			\vspace{0.2cm}
	\includegraphics[width=0.97\columnwidth]{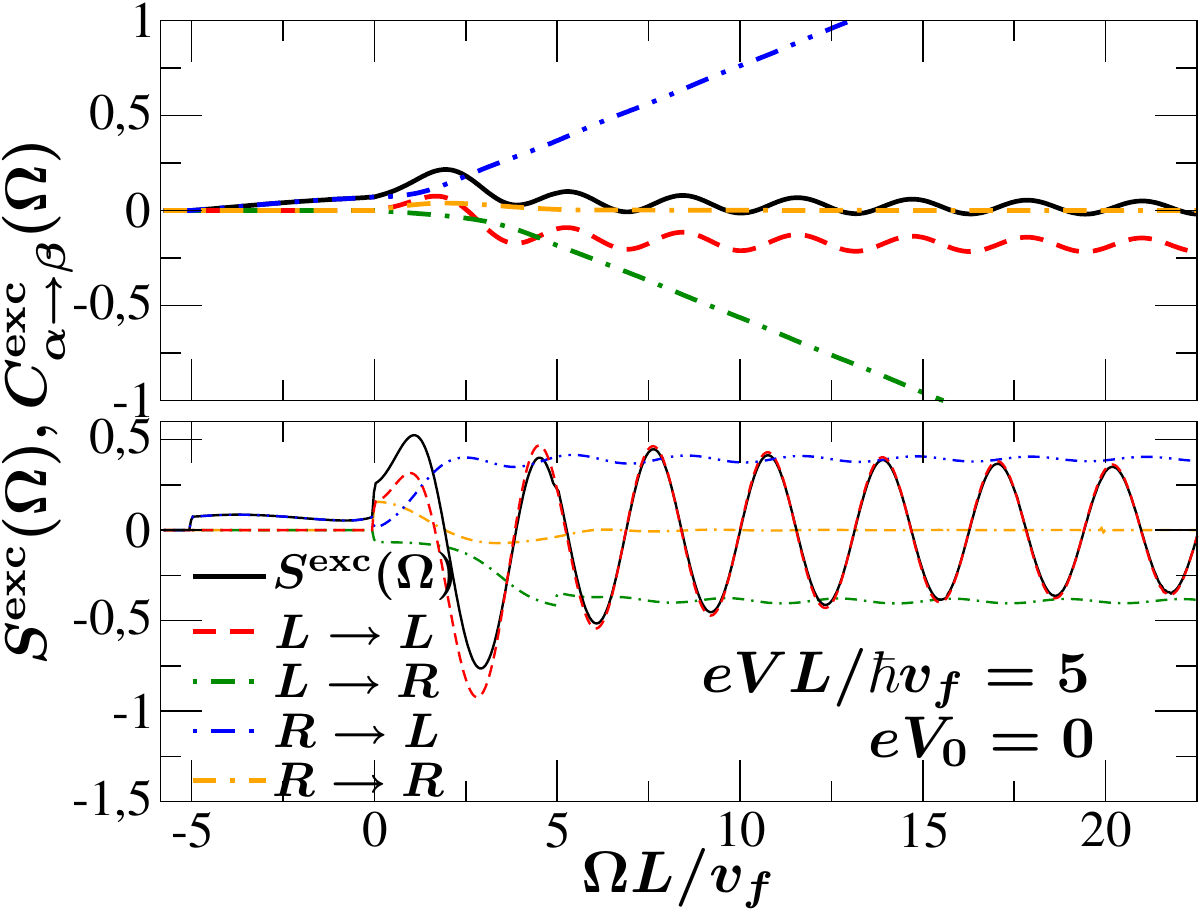}\\
	\vspace{0.6cm}
	\includegraphics[width=0.95\columnwidth]{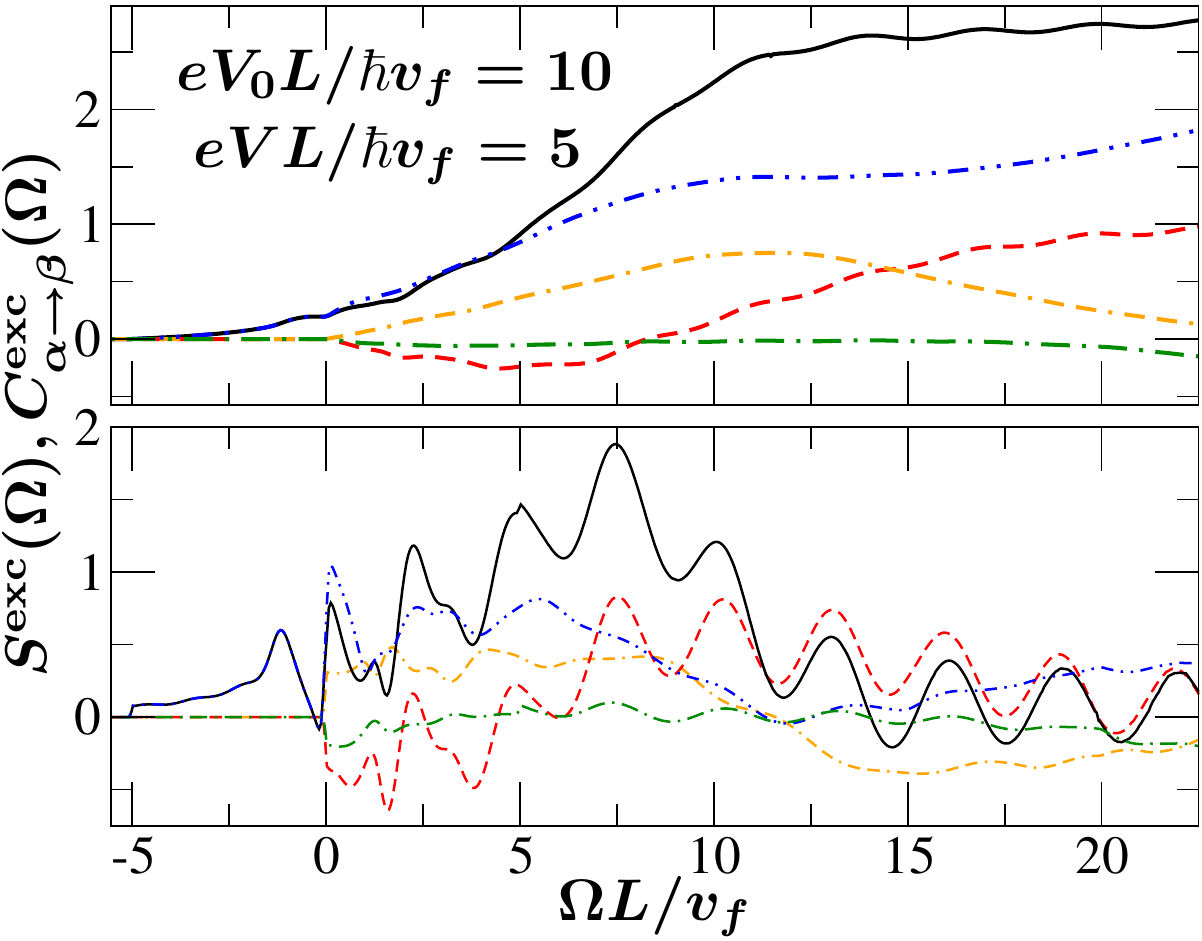}
	\vspace{0.01cm}
	\caption{(color online) Real parts of auto-correlation excess-noise spectrum in units of $2\pi\hbar/e^2$ (thick lines, upper panels) 
			and derivatives (thin lines, lower panels) with dc-bias symmetrically applied around 
			the Dirac point (top) and for finite $eV_0=2eV$ (bottom).
			By subtracting the noise at  zero dc-bias the divergent background 
			is removed. The structure and especially 
			the oscillatory behavior are coined by auto-terminal contributions 
			of Eq.(\ref{correlatorscross:LtoL}) related to the measurement terminal $L$. 
			The jump in the derivative of $C_{L \rightarrow R}(\Omega)$ is present because 
			this correlator does not contribute for frequencies  $\hbar \Omega <eV$.
			By applying an offset voltage $eV_0L/\hbar v_F=10$ a complicated 
			structure emerges, best visible in the derivatives.}
	\label{fig12}
\end{figure}
\subsection{Shot noise spectrum} 
In the regime $eV, \hbar \Omega, \hbar \omega \ll \hbar v_F/L$, the scattering matrix can be treated 
as energy-independent. Then, as for a single level quantum dot in the broad-band limit, 
asymmetric quantum noise as function of frequency is  the sum of four straight lines,
with kinks at $\hbar \Omega=0,\pm eV$.~\cite{GavishImry00,GavishImry02} 
For vanishing dc-bias we have
$C_{R \rightarrow L}^{\mathrm{}}(\Omega)=C_{L \rightarrow R}^{\mathrm{}}(\Omega)$ and 
$C_{R \rightarrow L}^{\mathrm{}}(\Omega)\approx C_{L \rightarrow R}^{\mathrm{}}(\Omega)$, as long as $ \Omega \ll v_F/L$. 
The richer regime, when  $eV, \hbar \Omega, \hbar \omega > \hbar v_F/L$, 
additionally exhibits strongly oscillating integrands.  
Those oscillations are purely due to propagating modes as it is also clear from interference patterns of 
the integrands in Figs.~\ref{fig5}-\ref{fig9}, regions II$_{a,b}$ and III$_{a,b,c}$.
In the special case of perpendicular incidence  ($q,\alpha(\epsilon)=0$) we have Klein tunneling, thus 
the frequency-dependence of the correlators is linear for this mode. 
Then  $C_{\alpha \rightarrow \beta}(\Omega)=0$ if $\alpha \ne \beta$ since $R(\epsilon)=0$.
Otherwise the $C_{\alpha \rightarrow \beta}(\Omega)$ mirror the interference patterns of the integrands.
So the noise spectrum (Fig.~\ref{fig11} solid, thick curve) shows oscillations on the scale of $L/\hbar v_F$ in the 
regime $eV, \hbar \Omega, \hbar \omega \gg \hbar v_F/L$, 
similar to the shot-noise at zero-frequency as a function of gate voltage~\cite{Tworzydlo06}. 
Although present in all four correlators, the oscillations show up in the noise spectrum
 mainly via $C^\mathrm{}_{L\rightarrow L}(\Omega)$ of the terminal where the fluctuating currents are probed. 
That is because the correlator itself as well as the amplitude of the oscillations are significantly larger than for other contributions. 
Therefore, in comparison to the absorption-branch (positive frequencies) the emission-branch of 
the spectrum (negative frequencies) shows only  small shot-noise. Indeed all correlators except $C_{R\rightarrow L}(\Omega)$ vanish when $\Omega \le 0$
since the energy for the emission of a photon has to be provided by the voltage source. 
Especially the contribution  dominant at positive frequencies vanishes: 
$C^\mathrm{}_{L\rightarrow L}(\Omega)=0$ if $\hbar \Omega \le 0$.\\

We are considering the limit $k_BT=0$ where the correlators integration windows are exactly determined by the chemical potentials. 
At finite temperature this so-defined onsets of the four contributions as a function of frequency 
are smeared out by the broadening of the Fermi-functions.
Clearly a gate voltage does not affect these onsets since it does not enter in the Fermi functions of the leads, 
but it still changes the transmission function resulting in a modified spectrum. 
\begin{figure}[tb]
	\centering
	\vspace{0.0cm}
		 \includegraphics[width=0.97\columnwidth]{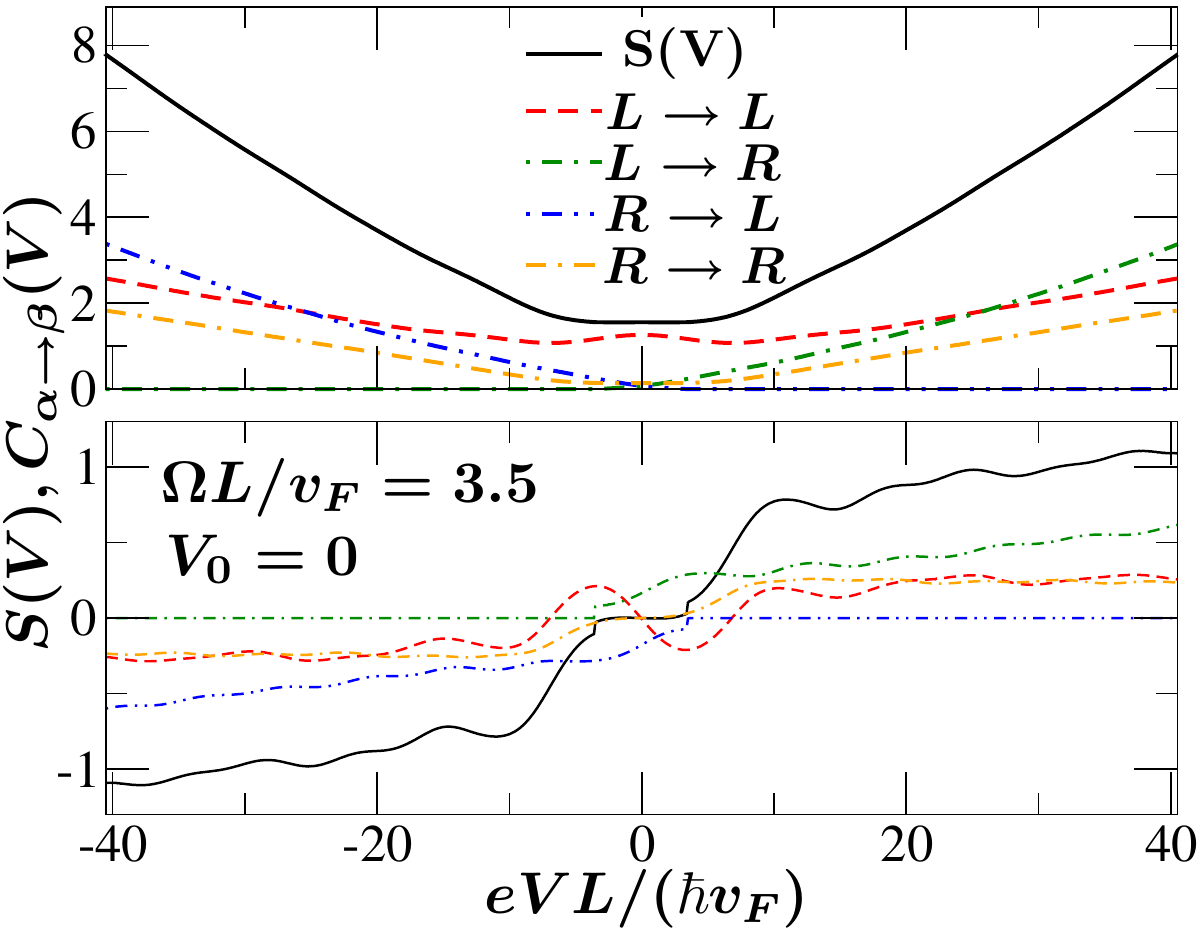}\\
				\vspace{0.0cm}
		 	\includegraphics[width=0.965\columnwidth]{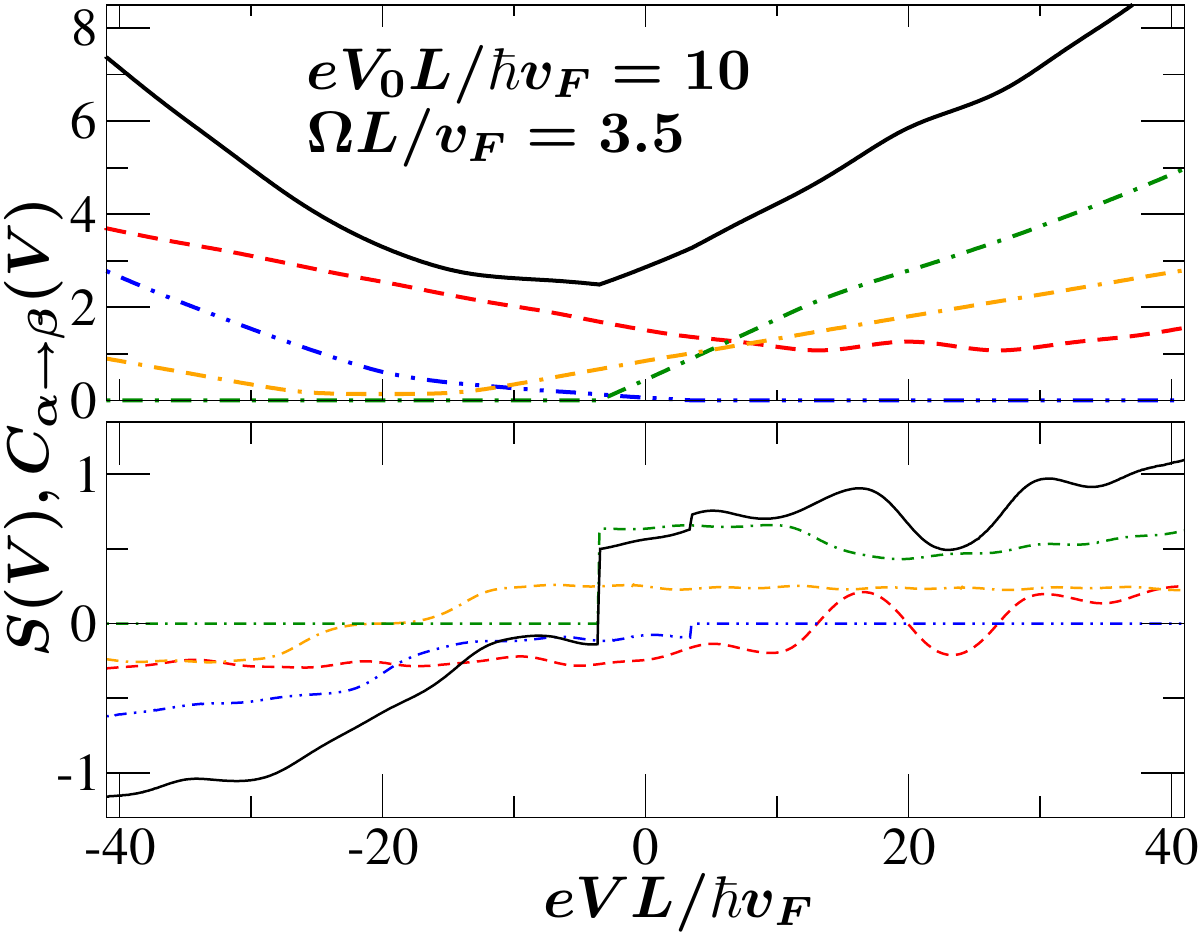}
	\caption{(color online) Real parts of auto-correlation current-current 
			fluctuations in units of $2\pi\hbar/e^2$ as a function of dc-bias for fixed frequency 
			$\hbar \Omega$ (thick lines, upper panels). We compare the symmetric setup without 
			dc-bias offset (top) and when $eV_0L/\hbar v_F=10$ (bottom). 
			Thin lines (lower panels) are used for the derivatives with respect to voltage.
			Top: Due to symmetrically applied bias voltage the noise and the auto-terminal contributions 
			are symmetric in the voltage dependence and 
			$\left.S_{\alpha \rightarrow \beta}(\Omega)\right|_{V}=\left.S_{\beta \rightarrow \alpha}(\Omega)\right|_{-V}$ if $\alpha\ne \beta$.
			Bottom: By applying an offset voltage we are breaking the setups symmetry. 
			Auto-terminal terms are then symmetric with respect to $eV=\pm 2eV_0$ while the summed up noise is asymmetric. }
	\label{fig13}
\end{figure}
Those limits of energy-integration, as well as their position relative to region III$_b$, result in features 
in the noise spectra besides the discussed oscillations.
In order to clarify the role of the Dirac Hamiltonian in comparison to 
the role of pure Fabry-P\'erot interferences, we compare results 
when the charge injection is only in the conduction or 
valence band by shifting the dc-bias voltages above
the Fermi energy of the graphene sheet via the offset 
voltage $V_0$ in $\mu_{L/R}=\pm eV/2 + eV_0$.
$C^\mathrm{}_{L\rightarrow R}(\Omega)$ can
never see the regime $-\hbar \Omega<\epsilon<0$ when $eV_0=0$, as in the upper plot of Fig.~\ref{fig11}. 
Thus the oscillations visible in the derivate 
have a well defined period over the whole spectrum on top of a
linearly increasing background.
When an offset voltage $eV_0= 2eV$ is applied, 
as done when calculating the spectra for the lower plot of Fig.~\ref{fig11}, 
$C_{R\rightarrow R}(\Omega)$ shows 
a  complicated frequency  dependence for small $\Omega$. 
Contribution $C_{L\rightarrow L}(\Omega)$ describes correlations of 
scattering states emanating from the left reservoir reflected back into 
the same reservoir. We will discuss this contribution now in detail: 
Special features for small frequency are due to the interplay of 
the integration boundaries with the various regions in Fig.~\ref{fig4}a) 
occurring in the integrands $(q,\epsilon)$-dependence of Fig.~\ref{fig7}a). 
Integration is  over all $q$-modes and from $\epsilon=-eV/2+eV_0-\hbar \Omega$ 
to $\epsilon=eV_0-eV/2$. When $eV_0=0,eV=0$ this corresponds 
to $-\hbar\Omega < \epsilon <0$, regions III$_b$ and partly II$_{a,b}$ 
of Fig.~\ref{fig4}a). Now at finite $eV,eV_0$ as in Fig.~\ref{fig11}, 
the integration window can include region III$_b$ completely, partly, or not at all, 
resulting in variations of the spectrum. At small $\hbar \Omega$, 
features in the integrands interference patterns have stronger impact. 
This can be seen from strongly non-harmonic features of the noise spectrum, 
e.g. in $C_{L\rightarrow L}(\Omega)$
and $C_{R\rightarrow L}(\Omega)$ for $eV_0=2eV$.
For large frequencies averaging leads to nearly harmonic oscillations 
on top of the increasing background. 
 With the chosen parameters the distance 
of the chemical potential $\mu_L$ to the charge-neutrality point is given by 
$e(-V/2+V_0) L/(\hbar v_F)=7.5$. Around the corresponding frequency the oscillatory behavior 
of the spectrum is modified and flattened due to a reduced fraction of propagating modes.
Raising the frequency further increases this fraction again and oscillations 
are roughly harmonic with period 
$\pi L/\hbar v_F$, best visible in the derivatives $dC_{L\rightarrow L}(\Omega)/d\Omega$ of Fig.~\ref{fig11}. 
That is also the point where the lower bound of energy integration starts 
to include the special interference pattern of the integrands around 
the energy interval $-\Omega< \epsilon < 0$, region III$_b$. 
$C_{R\rightarrow R}(\Omega)$ is not influenced 
by the measurement terminal itself, but probes transmission probabilities 
via scattering events which are related to the right terminal only. 
An analogous behavior of the spectrum as before is found, 
this time with a distance $e(V/2+V_0) L/(\hbar v_F)=12.5$ of the 
lower integration boundary to the charge neutrality point when 
$\hbar \Omega=0$. Now increasing frequency is going along with 
a decreasing slope of the derivative with respect to frequency 
until the Dirac point is reached. 
There the slope increases again since more open channels become available. 
The same interpretation also explains features in the interval $\hbar \Omega < eV$ 
of the auto-terminal correlators shown in Fig.~\ref{fig11}, when $V_0=0$. 
E.g. the spectrum of the correlator Eq.(\ref{correlators:RtoR}), with initial and final state 
in the right lead, exhibits a reducing slope until $\hbar \Omega=eV/2$ 
from where on the oscillations have a well defined period. 
The $dC_{R\rightarrow R}(\Omega)/d\Omega$ curve has a maximal slope at 
$\hbar \Omega=eV$ when positive and negative energies with same magnitude 
are present. For higher frequencies oscillations have again a well-defined phase.\\
We also study the excess noise at finite frequencies: 
$S_{\mathrm{exc}}(\Omega,\omega):=\left. S(\Omega,\omega)\right|_{eV}-\left. S(\Omega,\omega)\right|_{eV=0}$. 
Subtracting the noise at zero bias-voltage removes the divergent contributions 
from the noise spectrum. Then oscillating features due to bias-voltages are 
more obvious since they are now also prominent in the noise spectra of Fig.~\ref{fig12}, 
not only in derivatives. 
When $eV_0=0$ the excess noise (thick, black, solid curve) is purely positive for 
$\hbar \Omega \ll eV$ while for $\hbar \Omega > eV$ it is oscillating around zero, 
because then cross-terminal contributions $C_{\alpha \rightarrow \beta}(\Omega)$ cancel each other
up to a constant offset acquired at small $\Omega$. 
This offset is compensated by the $L\rightarrow L$
contribution. Oscillations of this contribution have again a considerable impact on the excess noise spectrum. 
In the lower plot of Fig.~\ref{fig12} the offset voltage is fixed to $eV_0=2eV$. 
For low frequencies $\hbar \Omega < eV$, complicated oscillations occur in all contributions
to excess noise and are accompanied by a strongly increasing slope up to frequencies $\hbar \Omega>eV_0 +eV/2$.   
As for the noise itself, the frequency of the oscillations is determined by $\hbar\Omega_Z=2eV$ and equals the frequency 
expected from the Zitterbewegung of relativistic Dirac fermions~\cite{Katsnelson06}. 
This frequency corresponds to a period of $T=\pi$ in our plots. 
It would be interesting to test experimentally if those much more pronounced oscillation, 
compared to the overall shot-noise, can be detected in spite of randomization effects 
of imperfections on the quasi-particles path lengths.
In summary, i) the impact of the Dirac Hamiltonian on the frequency-dependence of auto-terminal current fluctuations
leads to peculiar oscillation for energies in the vicinity of the Dirac point as an interplay of Klein tunneling, 
phase-jumps in the correlators and their energy-integration limits. And ii) oscillations due to the FP setup have a constant 
phase for high energies when propagating modes are dominant. Then $dS^{\mathrm{exc}}(\Omega)/d\Omega$ oscillates 
between positive and negative values with a period as it is expected from the effect of Zitterbewegung. 
\subsection{Dc-bias dependence at finite frequency}
\begin{figure}[tbp]
	\centering
			\vspace{0.6cm}
		\includegraphics[width=0.95\columnwidth]{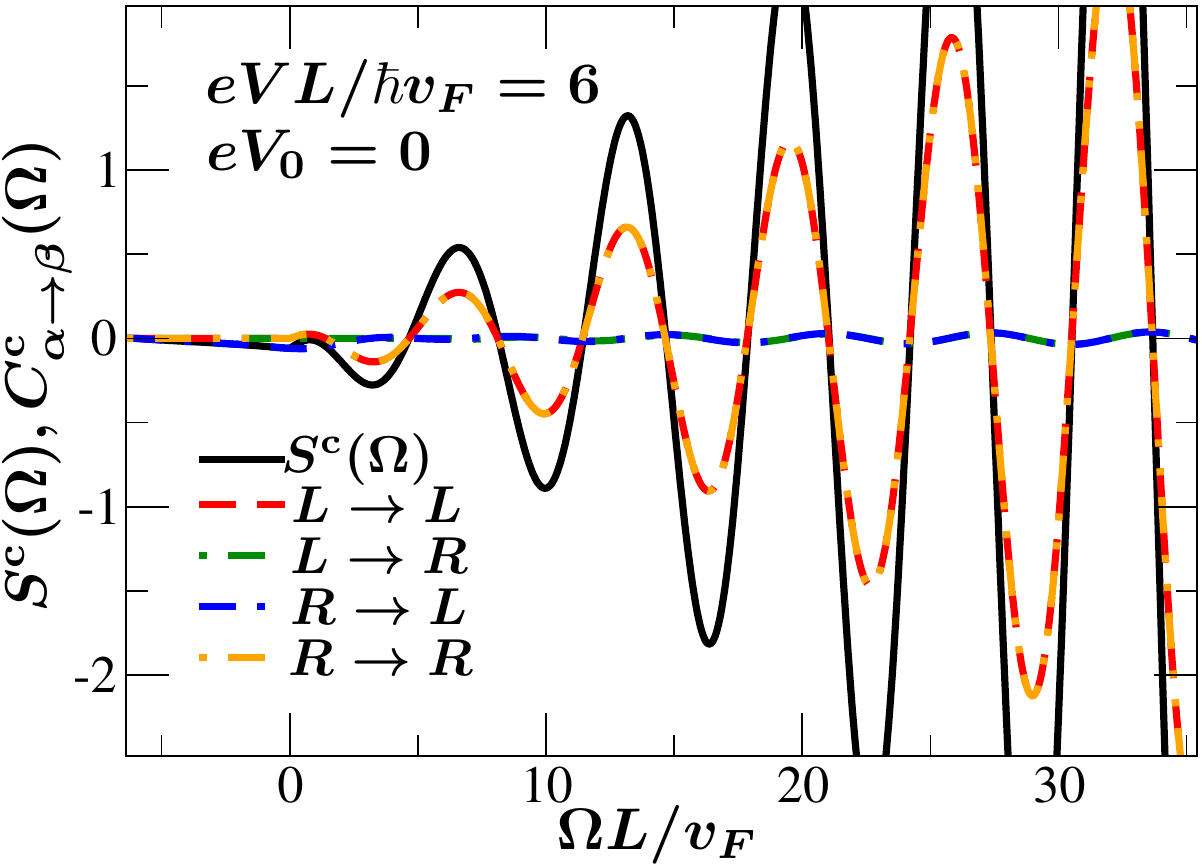}\\
		\vspace{1.0cm}
		\includegraphics[width=0.95\columnwidth]{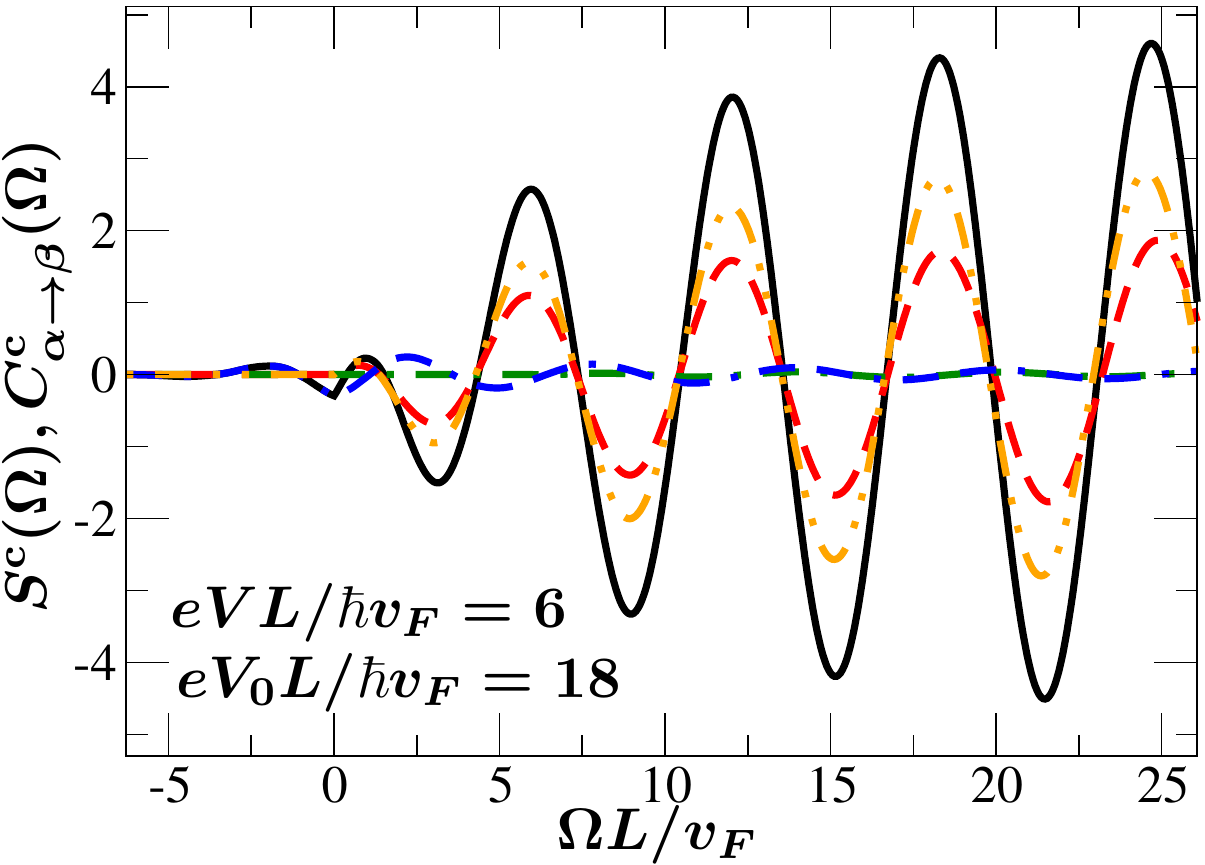}
	\caption{(color online) Real parts of cross-correlation spectrum in units of $2\pi\hbar/e^2$
			when $eV_0=0$ (top) and $eV_0L/\hbar v_F=3eVL/\hbar v_F=18$ (bottom).
			Without an offset voltage ($eV_0=0$) auto-terminal contributions are identical, 
			as well as cross-terminal ones at large frequencies $\hbar \Omega  \gg eV$. 
			At finite $eV_0$ the asymmetric bias voltage is reflected in the frequency-dependence of the 
			auto-terminal correlators by their different heights and the shift of the oscillations maxima.
			}
	\label{fig14}
\end{figure}
Analogous to the spectrum, the dc-bias dependence for fixed frequency is featureless in the regime 
$eV, \hbar \Omega, \hbar \omega \ll \hbar v_F/L$, 
except the pronounced onsets of the four correlators. 
This is not surprising when looking at the derivatives with respect to voltage:
\begin{subequations}
	\label{derivativesV}
\begin{align}
& \frac{dC_{L \rightarrow L}}{dV}=\frac{e^2 \Theta(\Omega) }{4\pi \hbar} \underset{-\infty}{\overset{\infty}\int}  dq \left[\left|1-r^*(-eV/2)r(\hbar\Omega-eV/2) \right|^2  \right.\nonumber\\ 
 &\left. -\left|1-r^*(-eV/2-\hbar \Omega)r(-eV/2) \right|^2 \right]  	\label{derivativesV:LtoL} \\
	&\frac{dC_{R \rightarrow R}}{dV}= \frac{e^2 \Theta(\Omega) }{4\pi \hbar}\underset{-\infty}{\overset{\infty}\int}  dq\nonumber\\
	&\left[T(eV/2)T(eV/2 +\hbar \Omega) -T(eV/2-\hbar\Omega)T(eV/2)\right]\label{derivativesV:RtoR}\\
 &\frac{dC_{L \rightarrow R}}{dV}=\frac{e^2 \Theta(\Omega-eV) }{4\pi \hbar}\underset{-\infty}{\overset{\infty}\int}  dq\nonumber\\ &\left[ T(\hbar\Omega-eV/2)R(-eV/2)-T(eV/2)R(eV/2-\hbar\Omega) \right]	\label{derivativesV:LtoR}\\
	& \frac{dC_{R \rightarrow L}}{dV}=\frac{e^2 \Theta(\Omega+eV) }{2\pi \hbar} \underset{-\infty}{\overset{\infty}\int}  dq\nonumber\\ &\left[  T(eV/2)R(eV/2+\hbar\Omega) -T(-eV/2-\hbar\Omega) R(-eV/2)\right]	\label{derivativesV:RtoL}
\end{align}
\end{subequations}
Scattering amplitudes are roughly constant for a given q-mode in this regime, 
then correlators are straight lines as a function of dc-bias voltage. 
E.g. a special situation that could exhibit interesting physics is when some derivatives are zero. 
But this is, due to symmetry arguments, only possible at $eV=0,\pm \hbar \Omega$, 
proofing a zero slope of the correlators at their onsets but revealing no additional effect.
By this means, as in the shot-noise spectrum, the dependence on the bias voltage 
reveals again the onsets of the four correlators. Since we  have chosen 
positive $\hbar \Omega$, the auto-terminal contributions  are non-zero over the whole bias range.
As before, cross-terminal ones vanish if no energy is provided by the voltage source: 
$C_{L \rightarrow R}\ne 0$ if $eV>-\hbar \Omega$
and $C_{R \rightarrow L}\ne 0$ if $eV<\hbar \Omega $. 
As it is clear from the bottom plot of Fig.~\ref{fig13}, 
the oscillations of the components are not in phase, 
thus adding up to complicated oscillations in $S_{LL}(\Omega)$. 
But, as mentioned in the beginning, we doubt this could be a  measurable effect. 
The shot-noise and the auto-terminal correlators are symmetric in the voltage dependence if $V_0=0$, 
whereas the cross terminal ones obey $C_{\alpha \rightarrow \beta}(\Omega,V)=C_{\beta \rightarrow \alpha}(\Omega,-V)$. 
Here the charge-neutrality point and the width of the region III$_b$ are revealed 
as a minima in the slope of the correlator $C_{L \rightarrow L}(\Omega)$  at $eV=\pm 2 \hbar \Omega$ 
and in the change of sign in $dC_{R \rightarrow R}(\Omega)/dV$ at $eV=0$. 
\section{Cross-correlation noise \label{sec:crossnoise}} 
\begin{figure}[tbp]
	\centering
	\vspace{0.6cm}
		\includegraphics[width=0.95\columnwidth]{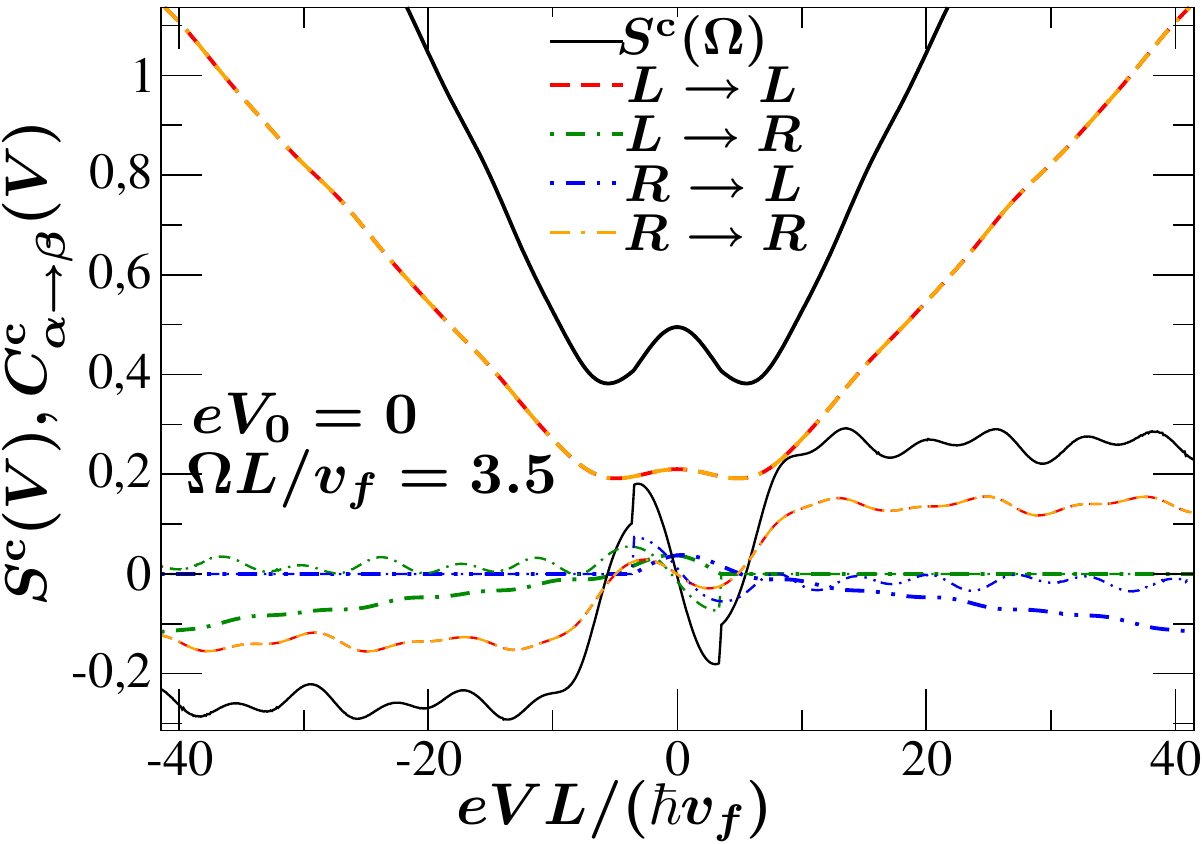}
	\caption{(color online) Real parts of current-current cross-correlations 
			in units of $2\pi\hbar/e^2$, as function of symmetrically 
			applied dc-bias voltage and for fixed frequency. 
			Jumps in the derivatives at  $\pm \Omega L/v_F$ are 
			due to the onsets of the cross-terminal contributions.} 
	\label{fig15}
\end{figure}

The explicit expressions of Eq.(\ref{correlatorscross}) for the cross-correlation current noise 
spectrum of Fig.~\ref{fig14} can be extracted from the general expression Eq.(\ref{noisedef}) in the 
same way as we did when deriving Eq.(\ref{correlators}). 
From Figs. ~\ref{fig8} a) and b) it is also clear that the spectrum of auto-terminal 
correlators are oscillating as a function of $\Omega$ with larger amplitude than cross-terminal ones, 
since they show an alternating behavior between positive and negative integrands. 
Dependence on $q$ in the relevant frequency range is weak, as shown in Fig.~\ref{fig8} a). 
Contrary, cross-terminal contributions as in Fig.~\ref{fig8} 
b) show features with an alternating sign along both variables, $\Omega$ and $q$. 
Thus, integration along y-momentum leads to averaging and therefore 
significantly smaller oscillation amplitudes occur. 
As discussed for the excess noise of the auto-correlation noise spectral function, 
we find the oscillations have a frequency $\hbar \Omega_Z=2eV$ what is tantamount to a period $T=2\pi$ in the plots.
Complex conjugation corresponds to time-reversed states. 
Again, as the product of scattering matrices of the integrands in Eq.(\ref{noisedef}) suggests, 
it is probing transmission- and reflection amplitudes 
of electron-hole pairs separated by an energy quanta $\hbar \Omega$. 
So, for cross-terminal noise not only the reflection but also the complex 
transmission amplitude is essential even without ac-bias voltages. 
Again it would be interesting to test if the resulting oscillations could be detected 
in the challenging task of a finite-frequency cross-correlations experiment. 
Analyzing the integrands reveals the symmetry 
$C^{\mathrm{c}}_{\alpha \rightarrow \alpha}(\Omega)=C^{\mathrm{c}}_{\beta \rightarrow \beta}(\Omega)$ if 
$\mu_L=-\mu_R$ as we show in Fig.~\ref{fig15}. This symmetry is distorted by applying an offset voltage $V_0$.
The spectrum of the correlator $C^{\mathrm{c}}_{L\rightarrow L}(\Omega)$ shows a shift of the maxima and minima of 
the oscillations with respect to  $C^{\mathrm{c}}_{R \rightarrow R}(\Omega)$ for finite $V_0$. 
This shift is due to the fact that the distance between neighboring maxima
of the integrand is not constant when varying $\hbar \Omega$ at given q-mode 
(see the bending of the maxima towards higher frequencies for larger $q$ in the integrands, e.g. Fig.~\ref{fig6}). 
Derivatives of the correlators $C^{\mathrm{c}}_{\alpha \rightarrow \beta}(\Omega)$ 
with respect to voltage show a sequence of pairs of different maxima. 
This observation is traced down to the same origin as above, 
and so the appearance of peculiar oscillations in the 
summed up cross-correlation shot-noise $S_{LR}(\Omega)$ is explained. 
At $\hbar \Omega=0$ current conservation and the unitarity of the s-matrix 
require $S_{LR}(\Omega)=-S_{LL}(\Omega)$. 
Therefore the correlator described by Eq.~\ref{correlatorscross:RtoL} is negative.
\section{Finite-frequency noise at ac-bias \label{sec:finitefreq}} 
By applying an ac-bias voltage at the leads one can inject
charge-carriers at positive and negative energies of the Dirac cone
without applying a dc-voltage. Analogous to the minimal conductivity,
in the non-driven case going along with a maximal Fano factor, the
shot-noise at zero frequency but finite ac-bias $S_{\alpha
  \alpha}(\Omega=0;\omega)$ mirrors the behavior of the conductivity
in Fig.~\ref{fig3}. 
\begin{figure}[tbp]
	\centering
	\vspace{0.0cm}
		\includegraphics[width=0.95\columnwidth]{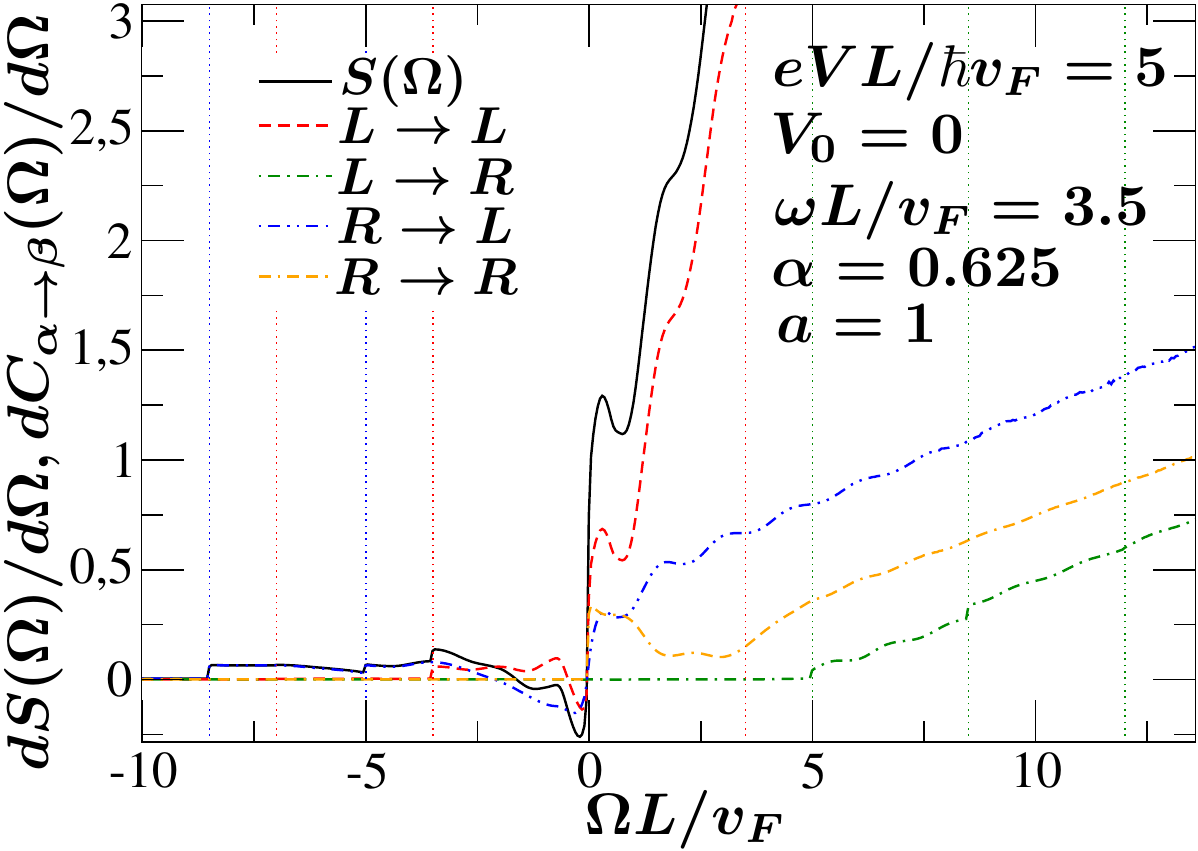}
	\caption{(color online) Derivatives of current-current correlations real parts 
			in units of $2\pi\hbar/e^2$ with respect to frequency  
			as function of frequency. We have chosen a symmetrically 
			applied dc-bias voltage with additional harmonic ac-driving 
			($\omega L/v_F=4$, $\alpha=0.5$ and $a=1$) in lead $L$. 
			Dashed vertical lines mark the step-positions, coloring specifies the correlator which 
			shows the step at corresponding $\Omega L/v_F$.} 
	\label{fig16}
\end{figure}
 The noise spectrum Fig.~\ref{fig16} for the driven setup ($a=1$) is 
similar to the one without driving but with additional steps in the
derivatives. For arbitrary ac-bias these steps can appear at frequencies 
$\hbar \Omega=\left(\mu_{\alpha}-\mu_{\beta}\right)\pm n \hbar \omega$ 
due to the onset of higher-order PAT events. Since we set $a=1$ in Fig.~\ref{fig16}, 
the correlator with states $R\rightarrow R$ shows no ac-induced steps in the derivative.
But when $|a|\ne 1$ all integrands (Fig.~\ref{fig9}) are not given in terms of probabilities
and can take negative values as mentioned in section \ref{sec:qualitative}. 
For the shot-noise spectrum there are then two possible sources of contributions that could reduce noise: 
Either a correlators integrand or the the product of Besselfunctions is negative.
When the driving voltage is applied symmetrically ($a=0$) more PAT-induced steps 
in the derivatives of the noise spectrum are visible and finite contributions 
at negative $\Omega$ are possible for all correlators.  
As proposed by Trauzettel et. al.,~\cite{Trauzettel07} a time-dependent
voltage could be used to induce interference between states in
particle- and hole-like parts of the Dirac spectrum. This should
correspond to Zitterbewegung like in relativistic quantum mechanics,
but we are not aware of any unique feature caused by Zitterbewegung
that can be distinguished from other oscillations, especially of Fabry-P\'erot
nature.
\section{Conclusions \label{sec:conclude}} 
We have analyzed conductivity and non-symmetrized finite-frequency current-current correlations 
for a Fabry-P\'erot graphene structure.
Oscillations on the intrinsic energy scale $L/\hbar v_F$ are still present in the finite frequency noise. 
Emission spectra are diverging for large frequencies, 
whereas the absorption branch of the spectrum has to vanish at $\hbar \Omega=-eV$. 
As expected from the integrands, the current-noise also diverges for voltages $|eV| \gg \hbar v_F/L $. 
Since the onset of the different noise contributions is defined by the four 
possible combinations of the chemical potentials, the  noise built by 
all correlators consists of contributions oscillating with the same period but 
different phases. Although dominated by  $C_{L\rightarrow L}(\Omega)$ when 
correlating the currents in terminal $L$ at large frequencies, this interplay 
is revealed in the spectra and voltage dependence of all correlators. 
Each contribution can show peculiar oscillations at low enough frequencies 
or voltages. In this regime features in the integrands 
$(q,\epsilon)$-dependence  can have a prominent impact whereas they tend to be averaged out at large frequencies.  
Another aspect is the appearance 
of a special region showing phase jumps
in the energy dependence of the integrands when $\hbar \Omega \le 2eV$. 
This interplay of the Dirac spectrum and the Fabry-P\'erot physics~\cite{Liang01,Herrmann07}   
can be probed purely by applying an appropriate combination of dc-bias and offset voltage $V_0=V/2$, 
thus connecting electron- and hole part of the Dirac spectrum symmetrically when $eV=2 \hbar \Omega$. 
The way the scattering amplitudes are combined in this approach spoils the 
clear picture in terms of transmission- and reflection probabilities. Instead, in the dc-limit it gives 
rise to the interpretation of the $L \rightarrow L$ contribution in terms of jumps in the scattering-phase
 between time-reversed electron-hole states separated by the photon energy $\hbar \Omega$. 
In the same way the complex correlators for cross-correlation noise or for the driven setup 
exhibit phase jumps and can not be written in terms of probabilities. 
Complex contributions of the scattering matrices lead to large oscillations between positive and 
negative values of cross-correlation noise or in the derivatives 
with respect to frequency of the auto-terminal noise spectral function. 
These oscillations have a frequency of $\hbar\Omega_Z=2eV$, what corresponds to a period of $T=2\pi$ in our plots. 
This frequency corresponds to the Zitterbewegung frequency as it is known for relativistic Dirac fermions. 
Again, strongly non-harmonic features can occur when the transition between different regimes 
is probed, especially when region III$_b$ around the Dirac point comes into play. 
Additional ac-bias complicates the picture because 
combinations of $q, \hbar \Omega, m\hbar \omega$ define additional phase jumps, 
onsets of the correlators and therefore steps in the noise when higher-order PAT events occur. 
Then the special role of the complex reflection and transmission amplitudes 
is essential for all possible correlators.
\acknowledgments
We would like to acknowledge the financial
support by the DFG (Grant No. SFB 767) and 
thank B. Trauzettel for validating and sharing 
corrections to reference~\cite{Trauzettel07}.
\appendix
\section{Noise formulas \label{app:noise}}
The non-symmetrized noise spectrum under harmonic ac-driving without interactions is determined by 
\begin{widetext}
\begin{align}
 &S_{\alpha \beta}(\Omega,\omega) = \left(\frac{e^2}{2 \pi \hbar}\right) \int d\epsilon \underset{\gamma \delta, l k m}\sum
J_l\left(\frac{eV_{\mathrm{ac},\gamma} }{\hbar \omega} \right) J_k\left(\frac{eV_{\mathrm{ac},\delta} }{\hbar \omega} \right) J_{m+k-l}\left(\frac{eV_{\mathrm{ac},\delta} }{\hbar \omega} \right) J_m\left(\frac{eV_{\mathrm{ac},\gamma} }{\hbar \omega} \right) \nonumber\\
&  Tr\left[{A}_{\gamma \delta}(\alpha, \epsilon, \epsilon + \hbar \Omega) 
{A}_{\delta \gamma}(\beta, \epsilon + \hbar \Omega + (m-l) \hbar \omega, \epsilon+ (m-l) \hbar \omega, ) \right]
f_{{\gamma}}(\epsilon-l \hbar \omega) \left(1-f_{{\delta}}(\epsilon +\hbar \Omega -k \hbar \omega)  \right) \, .
\label{thenoise}
\end{align}
\end{widetext}
For the Fermi distribution function in lead $\gamma$ we use the shorthand 
$f_{\gamma}(\epsilon)=1/(\exp\left[(\epsilon-\mu_{\gamma})/k_BT\right]+1)$. 
In the limit of $k_BT=0$ the distribution functions in the leads are given 
by Heaviside-Theta functions $\Theta(\mu_{\gamma}-\epsilon)$ that define the integration intervals. 
Explicitly writing down the expression above for chosen $\alpha, \beta$ then leads 
to the four possible contributions to auto-correlation noise of Eq.~(\ref{correlators}) 
and to cross-correlation noise of Eq.~(\ref{correlatorscross}) via
summation over reservoir indices $\gamma,\delta=L,R$. 
\section{Boundary conditions \label{app:bounds}}
As in~\cite{Tworzydlo06} we confine the charge-carriers along the $y$-directions 
by infinite mass boundaries that diverges at the edges $y=0,W$. 
This corresponds to the boundary conditions~\cite{Berry87}
\begin{align}
\left. \hat{\Psi}_1\right|_{y=0}= \left.\hat{\Psi}_2\right|_{y=0} \qquad \left. \hat{\Psi}_1\right|_{y=W}= -\left.\hat{\Psi}_2\right|_{y=W}. 	 
\end{align}
Unlike the procedure for the Schr\"odinger equation, in
graphene one only has to match the wave function itself and no
constraint is given for the derivatives.
Now exploiting the boundary conditions along the $x$-direction, 
transmission- and reflection amplitudes are fully determined by 
these constraints of the field operators at the Fermi levels. 
Therefor it is sufficient to match the wave functions at $x=0,L$ without ac-driving. 
A plain wave ansatz to solve the Dirac equation~(\ref{diraceq}) for an electron 
incident from the left ($x<0$) with energy $\epsilon$ is given by 
\begin{equation}
	\Psi({\mathbf x})=\left\{
\begin{array}{ccc}
	\Psi_{0,+}^{\kappa q}+ r(\epsilon)\, \Psi_{0,-}^{\kappa q} & {\mathrm{ if }} & x<0 \\ 
	 a(\epsilon)\, \Psi_{0,+}^{k q}	+b(\epsilon)  \,  \Psi_{0,-}^{k q}& {\mathrm{ if }} & 0<x<L \\ 
	t(\epsilon)   \Psi_{0,+}^{\kappa' q} e^{-i\kappa' L} & {\mathrm{ if }} & x>L
\end{array}
\right.
\end{equation}
where  $\kappa(\epsilon),\kappa'(\epsilon)$  are the complex wave vectors 
in the reservoirs and and $k(\epsilon)$ the one in the sandwiched graphene strip. 
Matching conditions at boundaries (continuity at $x=0,L$) combined with high doping 
in the reservoirs lead to the following set of coupled, complex equations:
\begin{align}
	\frac{1 + r(\epsilon)}{\sqrt{2}}  &=	 a(\epsilon) \frac{e^{- i \alpha(\epsilon)/2}}{\sqrt{\cos(\alpha(\epsilon))}} +   b(\epsilon) \frac{e^{ i \alpha(\epsilon)}}{\sqrt{\cos(\alpha(\epsilon))}}\\
	\frac{1 - r(\epsilon)}{\sqrt{2} } &=	 a(\epsilon) \frac{e^{ i \alpha(\epsilon)/2}}{\sqrt{\cos(\alpha(\epsilon))}} -  b(\epsilon) \frac{e^{- i \alpha(\epsilon)}}{\sqrt{\cos(\alpha(\epsilon))}}\\
	\frac{t(\epsilon)}{\sqrt{2} } &=	 a(\epsilon) \frac{e^{ i k(\epsilon) L}e^{- i \alpha(\epsilon)/2}}{\sqrt{\cos(\alpha)}} +   b(\epsilon) \frac{e^{- i k(\epsilon) L}e^{i \alpha(\epsilon)}}{\sqrt{\cos(\alpha)}}\\
	\frac{t(\epsilon)}{\sqrt{2}}  &=	 a(\epsilon) \frac{e^{ i k(\epsilon) L}e^{ i \alpha(\epsilon)/2}}{\sqrt{\cos(\alpha)}} -  b(\epsilon) \frac{e^{- i k(\epsilon) L}e^{- i \alpha(\epsilon)}}{\sqrt{\cos(\alpha(\epsilon))}}
\end{align}
The solution is straightforward and determined by the four complex 
coefficients $a(\epsilon),b(\epsilon),t(\epsilon),r(\epsilon)$. 
The transmission $t(\epsilon)$ and reflection $r(\epsilon)$ amplitudes define the s-matrix 
and thus the current and noise of the scattering device at all voltages.
\begin{align}
	t(\epsilon)&= \frac{1}{\cos(k(\epsilon) L)-i \sec(\alpha(\epsilon) ) \sin(k(\epsilon) L)}\\
	r(\epsilon)&= \frac{1}{-\cot(k(\epsilon) L) \cot(\alpha(\epsilon) )+ i \text{ csc}(\alpha(\epsilon) )}\\
	a(\epsilon)&= \frac{\sqrt{2} \cos\left(\frac{\alpha(\epsilon) }{2}\right) \sqrt{\cos(\alpha(\epsilon) )}}{1+e^{2 i k(\epsilon) L} (-1 + \cos(\alpha(\epsilon) )) + \cos(\alpha(\epsilon) )}\\
	b(\epsilon)&= -\frac{e^{i k(\epsilon) L} \sqrt{\cos(\alpha(\epsilon) )} \sin\left(\frac{\alpha(\epsilon) }{2}\right)}{\sqrt{2} (i \cos(k(\epsilon) L) \cos(\alpha(\epsilon) ) + \sin(k(\epsilon) L))}
\end{align}
\section*{References}
\end{document}